\documentclass[aps,prd,reprint,twocolumn,superscriptaddress]{revtex4-2}
\usepackage[english]{babel}
\usepackage[T1]{fontenc}
\usepackage{amsmath}
\usepackage{amssymb}
\usepackage{graphicx}
\usepackage{color}
\usepackage{times}
\usepackage{euscript}
\usepackage{amsthm}
\usepackage{amsfonts}
\usepackage{epstopdf}
\usepackage{multirow,makecell,array}
\usepackage{mathtools}
\usepackage{float}
\usepackage[dvipsnames]{xcolor}

\newcommand{\ngamma}{n_{\gamma}}
\newcommand{\nbgamma}{\bar{n}_{\gamma}}

\begin{document}

\title{Helicity effects in the dynamically assisted Schwinger mechanism}

\author{A.~I.~Baksheev}
\affiliation{Department of Physics, Saint Petersburg State University, Universitetskaya Naberezhnaya 7/9, Saint Petersburg 199034, Russia}
\author{V.~A.~Bokhan}
\affiliation{Department of Physics, Saint Petersburg State University, Universitetskaya Naberezhnaya 7/9, Saint Petersburg 199034, Russia}
\author{A.~Kudlis}
\affiliation{Science Institute University of Iceland, Reykjavik, IS-107, Iceland}
\author{I.~A.~Aleksandrov}
\affiliation{Department of Physics, Saint Petersburg State University, Universitetskaya Naberezhnaya 7/9, Saint Petersburg 199034, Russia}

\begin{abstract}
We study vacuum electron-positron pair production in a spatially uniform bichromatic electric field within the corrected quantum-kinetic framework for fermions. The external background models the superposition of two counterpropagating circularly polarized laser pulses and combines a strong slowly varying component with a weak rapidly oscillating one. We analyze the weak-field multiphoton regime, the strong-field tunneling regime, and their combination corresponding to the dynamically assisted Schwinger effect. Our main focus is on helicity-resolved observables. We show that dynamical assistance enhances not only the total yield but also the helicity asymmetry: right- and left-handed electrons preferentially populate opposite momentum half-spaces. Most importantly, within the parameter range considered here, the ratio of the momentum distributions for opposite helicities is governed predominantly by the polar angle with respect to the propagation axis and depends only weakly on the momentum magnitude and azimuthal angle. The corresponding asymmetry becomes more pronounced as the weak-field frequency is increased. These results identify a clear helicity signature of the dynamically assisted Schwinger effect in rotating strong-field backgrounds and provide a compact characterization of the associated helicity-resolved spectra.
\end{abstract}

\maketitle

\section{Introduction}
\label{sec:intro}

Electron-positron pair creation from vacuum in a strong external electromagnetic field is one of the central nonperturbative predictions of strong-field quantum electrodynamics (QED). In the simplest idealized setting of a uniform static electric field, this phenomenon is described by the Sauter--Schwinger mechanism~\cite{sauter_1931,heisenberg_euler,schwinger_1951}. Although a direct experimental observation remains challenging because the relevant field strengths are extremely large, rapid progress in intense-laser physics continues to stimulate strong interest in this problem; for broader background and recent developments we refer to the reviews~\cite{xie_review_2017,gonoskov_2022,fedotov_review}. In realistic laser setups, the external background is inherently time dependent and often multiscale, which makes it especially important to identify field configurations that can both enhance the production probability and reveal robust observable signatures of the process.

A particularly important mechanism in this context is the dynamically assisted Schwinger effect, where a strong slowly varying field is supplemented by a weaker rapidly oscillating component~\cite{schuetzhold_prl_2008}. Physically, the fast mode effectively reduces the tunneling suppression and can strongly enhance pair creation while preserving the essentially nonperturbative character associated with the strong field. Since the original proposal, dynamical assistance and closely related scenarios have been explored in a broad variety of settings, including momentum distributions and pulse-shape dependence~\cite{orthaber_plb_2011,fey_prd_2012,nuriman_plb_2012,kohlfuerst_prd_2013,akal_prd_2014,hebenstreit_plb_2014,linder_prd_2015,otto_plb_2015,otto_prd_2015}, extensions beyond the spatially uniform approximation~\cite{aleksandrov_prd_2018,li_prd_2021}, and perturbative as well as worldline aspects of assistance~\cite{torgrimsson_jhep_2017,torgrimsson_prd_2019}. Related multiscale and temporally structured setups, including chirped, asymmetric, and relative-phase-sensitive configurations, were studied in Refs.~\cite{dunne_prd_2009,schneider_jhep_2016,taya_prr_2020,kohlfuerst_prr_2021,sitiwaldi_pra_2023,chen_prd_2024,aleksandrov_sevostyanov_2025,brass_arxiv_2025,jangir_prd_2026}. Taken together, these studies have established dynamical assistance and related multiscale field engineering as a broad framework for understanding how temporal structure reshapes nonperturbative vacuum decay, momentum spectra, and interference patterns.

In parallel, increasing attention has been devoted to spin-, chirality-, and helicity-sensitive observables in vacuum pair production. Early studies of polarized and rotating backgrounds already indicated that the spin properties of the produced particles can carry nontrivial information about the geometry of the external field~\cite{woellert_prd_2015,blinne_strobel_2016,li_prd_2017,huang_2019,olugh_prd_2019,kohlfuerst_prd_2019,li_prd_2019,olugh_plb_2020}. More recently, this direction has developed particularly rapidly. Momentum spirals, chirality-related effects, and spin asymmetries were analyzed in Refs.~\cite{hu_prd_2023,yu_prd_2023,hu_prd_2024}; helicity- and spin-resolved momentum distributions were investigated within scattering, kinetic, and Dirac--Heisenberg--Wigner approaches in Refs.~\cite{majczak_prd_2024,aleksandrov_kudlis_prdl_2024,jiang_arxiv_2025}; pair production in circularly polarized waves and in two-color rotating fields was further examined in Refs.~\cite{kohlfuerst_prdl_2024,chen_prd_2025}; and spatially asymmetric time-dependent backgrounds were explored in Ref.~\cite{bake_prd_2025}. At the same time, the theoretical description itself has been refined. In particular, the corrected quantum-kinetic formulation for spatially uniform electric fields of arbitrary polarization was established in Ref.~\cite{aleksandrov_kudlis_klochai}, where it was shown that the proper fermionic QKE system contains ten rather than twelve independent functions, improving upon earlier formulations~\cite{pervushin_skokov}. A recent independent derivation and comparison with the Dirac--Heisenberg--Wigner formalism was presented in Ref.~\cite{li_arxiv_2025}, while further aspects of the revised QKE framework, including renormalized observables, were analyzed in Ref.~\cite{aleksandrov_arxiv_2026}. In addition, Ref.~\cite{jiang_arxiv_2025} formulated a general spin-resolved momentum distribution within the Dirac--Heisenberg--Wigner approach and identified the helicity-resolved distribution as a special case, thereby clarifying the relation between general spin projections and the helicity observables relevant for rotating-field pair production. Taken together, these developments make it timely to revisit the dynamically assisted Schwinger effect with particular emphasis on helicity-sensitive observables.

In the present paper, we employ the corrected QKE framework~\cite{aleksandrov_kudlis_klochai,aleksandrov_arxiv_2026} to investigate vacuum pair production in a spatially uniform bichromatic electric field which models, in the dipole approximation, the superposition of two counterpropagating circularly polarized laser pulses. Our aim is to connect two active directions of current research: dynamical assistance and helicity-resolved strong-field observables. Within this unified setup, we analyze the weak-field multiphoton regime, the strong-field tunneling regime, and the dynamically assisted regime, and then focus on the helicity-resolved momentum distributions of the produced electrons.

The central new result of this study is that dynamical assistance enhances not only the total pair yield but also the helicity asymmetry. For circular polarization, right- and left-handed electrons preferentially populate opposite $q_z$ half-spaces. More importantly, we find that, within the parameter range considered here, the ratio of the momentum distributions for opposite helicities is governed predominantly by the polar angle with respect to the propagation axis and depends only weakly on the momentum magnitude and azimuthal angle. This simple angular organization is physically natural: in a rotating field, the propagation axis is the distinguished direction selected by the field angular momentum, so it is reasonable that the helicity imbalance is governed primarily by the orientation of the particle momentum relative to that axis. To the best of our knowledge, such a compact characterization of the helicity-resolved spectra has not been identified previously for the dynamically assisted bichromatic setup considered here. We identify this feature as the central new result of the present work and as a distinct helicity signature of the dynamically assisted Schwinger effect in rotating strong-field backgrounds.

The paper is organized as follows. In Sec.~\ref{sec:setup} we specify the external-field configuration and pulse parameters. In Sec.~\ref{sec:qke} we summarize the corrected quantum-kinetic formalism and define the helicity-resolved observables used in our analysis. In Sec.~\ref{sec:res} we present the numerical results. We first discuss the spin-summed momentum spectra in the weak-field, strong-field, and dynamically assisted regimes, and then analyze the helicity-resolved spectra, quantify the helicity asymmetry, and study its angular dependence. Technical details of the locally constant field approximation are collected in the Appendix. Section~\ref{sec:concl} contains our conclusions.

Throughout this paper, we assume $\hbar = c = 1$ and use $e<0$ for the electron charge.


\section{Setup}
\label{sec:setup}

We consider an external field described by the vector potential
\begin{equation}
\mathbf{A}(t) = \sum_{i=1,2} \frac{E_i/\omega_i}{\sqrt{1+\delta_i^2}} \, F(\omega_1 t)
\begin{pmatrix}
\cos(\omega_i t + \varphi_i) \\
\delta_i \sin(\omega_i t + \varphi_i) \\
0
\end{pmatrix},
\label{eq:field}
\end{equation}
where $E_i$ and $\omega_i$ denote the amplitude and frequency of the corresponding field component, with $i=1$ labeling the strong low-frequency mode and $i=2$ the weak high-frequency mode. The parameters $\varphi_i$ are the carrier-envelope phases, while $\delta_i$ determine the polarization: $\delta_i=0$ corresponds to linear polarization along the $x$ axis, whereas $|\delta_i|=1$ corresponds to circular polarization in the $xy$ plane, with the sign of $\delta_i$ fixing the handedness. We work in the temporal gauge, where the scalar potential vanishes, $A_0=0$. In the present study, both components are taken with a common Gaussian envelope,
\begin{equation}
F(\eta) = \exp\!\left(-\frac{\eta^2}{\sigma^2}\right),
\label{eq:F1}
\end{equation}
where $\sigma$ is a dimensionless parameter controlling the pulse duration.

Since the vector potential~\eqref{eq:field} is spatially uniform, the corresponding external electric field, $\mathbf{E}(t)=-\dot{\mathbf{A}}(t)$, depends only on time, while the magnetic field vanishes. Each field component may be viewed as the dipole approximation to two counterpropagating laser pulses with suitably chosen polarizations. Because the strong component ($i=1$) produces pairs predominantly near the maximum of the electric field, the dipole approximation is expected to be well justified in the regime of interest; see Ref.~\cite{tkachev_pra_2025} and references therein for a recent discussion. The weak component ($i=2$) can, in principle, generate spatially nonlocal effects~\cite{aleksandrov_prd_2018}, which are beyond the scope of the present study. In the present work, however, we restrict ourselves to the simplified spatially uniform setup, which is sufficient for identifying the qualitative helicity effects that are the main focus of the present analysis.

For later use, we introduce the Keldysh parameter for each field component, $\gamma_i = m\omega_i/|eE_i|$. In a dynamically assisted configuration, one typically has $\gamma_1 \ll 1$ for the strong slowly varying field and $\gamma_2 \gg 1$ for the weak rapidly oscillating one, while the mixed parameter $\gamma_{\mathrm{c}} \equiv m\omega_2/|eE_1|$ characterizes the assistance mechanism~\cite{schuetzhold_prl_2008}.


\section{Quantum kinetic equations}
\label{sec:qke}

As shown in Ref.~\cite{aleksandrov_kudlis_klochai}, to zeroth order in the radiative interaction the pair-production process is governed by the following system of ordinary differential equations for ten unknown functions $f$, $\mathbf{f}$, $\mathbf{u}$, and $\mathbf{v}$:
\begin{align}
\dot{f} &= -2 \boldsymbol{\mu}_2 \mathbf{u} \,, \label{eq:system_simple_f_s}\\
\dot{\mathbf{f}} &= 2 (\boldsymbol{\mu}_1 \times \mathbf{f}) - 2 (\boldsymbol{\mu}_2 \times \mathbf{v}) \,, \label{eq:system_simple_f_v} \\
\dot{\mathbf{u}} &= 2 (\boldsymbol{\mu}_1 \times \mathbf{u}) + \boldsymbol{\mu}_2 (2f - 1) + 2 \omega \mathbf{v} \,, \label{eq:system_simple_u_v} \\
\dot{\mathbf{v}} &= 2 (\boldsymbol{\mu}_1 \times \mathbf{v}) - 2 (\boldsymbol{\mu}_2 \times \mathbf{f}) - 2 \omega \mathbf{u} \,. \label{eq:system_simple_v_v}
\end{align}
Here
\begin{align}
\boldsymbol{\mu}_1 (\mathbf{p}, t) &= \frac{e}{2\omega (\omega+m)}\, [\mathbf{q} \times \mathbf{E}(t)] \,, \\
\boldsymbol{\mu}_2 (\mathbf{p}, t) &= \frac{e}{2\omega^2 (\omega+m)}\, \big\{ [\mathbf{q} \mathbf{E}(t)] \mathbf{q} - \omega (\omega+m) \mathbf{E}(t) \big\} \,,
\end{align}
where $\mathbf{q} \equiv \mathbf{q}(t) \equiv \mathbf{p}-e\mathbf{A}(t)$ and $\omega \equiv \sqrt{m^2 + \mathbf{q}^2}$. All QKE functions vanish at the initial time $t=t_\text{in}\to -\infty$. Although these functions are parametrized by $\mathbf{p}$, the quantity $\mathbf{q}$ represents the kinetic (gauge-invariant) momentum of the particles. Since the external vector potential vanishes at $t=t_\text{out}\to +\infty$ in our gauge choice, one has $\mathbf{q}(t_\text{out})=\mathbf{p}$. Thus, the canonical and kinetic momenta coincide at asymptotically large times, and $\mathbf{p}$ may be identified with the final kinetic momentum. In practical calculations, it is sufficient to choose $t_\text{out}=-t_\text{in}\gg \sigma/\omega_1$, so that the external field is effectively switched off at the integration boundaries. The evolved QKE functions at $t=t_\text{out}$ can then be used to extract the observable momentum distributions.

In particular, the helicity-resolved momentum distributions of electrons are given by~\cite{aleksandrov_kudlis_klochai,aleksandrov_kudlis_prdl_2024}
\begin{equation}
f^{(e^-\text{L/R})}(\mathbf{p}, t) = f(\mathbf{p}, t) \mp \frac{\mathbf{q} \mathbf{f}(\mathbf{p}, t)}{|\mathbf{q}|}.
\label{eq:el_LR}
\end{equation}
Evaluating these expressions at $t=t_\text{out}$, one obtains
\begin{equation}
\frac{(2\pi)^3}{V} \frac{dN^{(e^-\text{L/R})}_{\mathbf{p}}}{d^3 \mathbf{p}} = f^{(e^-\text{L/R})}(\mathbf{p}) \equiv f^{(e^-\text{L/R})}(\mathbf{p}, t_\text{out}).
\end{equation}
Here $V$ is the normalization volume. The distributions $f^{(e^-\text{L/R})}(\mathbf{p})$ do not exceed unity, in accordance with the Pauli exclusion principle~\cite{aleksandrov_kudlis_klochai}. The helicity-summed, or equivalently spin-summed, electron spectrum is
\begin{equation}
f^{(e^-)}(\mathbf{p}) = f^{(e^-\text{L})}(\mathbf{p}) + f^{(e^-\text{R})}(\mathbf{p}) = 2 f(\mathbf{p}, t_\text{out}).
\label{eq:f_spin-summed}
\end{equation}
The corresponding positron distributions are obtained from
\begin{equation}
f^{(e^+\text{L/R})}(\mathbf{p}, t) = f^{(e^-\text{R/L})}(-\mathbf{p}, t).
\label{eq:pos_LR}
\end{equation}
Within the Dirac--Heisenberg--Wigner formalism, the helicity-resolved momentum distribution may be viewed as a special case of a more general spin-resolved construction based on the covariant spin projection operator~\cite{jiang_arxiv_2025}.

In what follows, we solve the QKE system~\eqref{eq:system_simple_f_s}--\eqref{eq:system_simple_v_v} numerically and analyze both the spin-summed and helicity-resolved momentum distributions, $f^{(e^-)}(\mathbf{p})$ and $f^{(e^-\text{L/R})}(\mathbf{p})$.


\section{Numerical results} \label{sec:res}

We now present numerical results obtained by solving the QKE system~\eqref{eq:system_simple_f_s}--\eqref{eq:system_simple_v_v}. In Sec.~\ref{subsec:ch2_spectra} we briefly discuss the \emph{spin-summed} momentum spectra for the weak field, the strong field, and their combination,
and interpret the dominant spectral structures using semiclassical turning-point arguments. In Sec.~\ref{subsec:helicity_asym} we analyze \emph{helicity-resolved} spectra, quantify the helicity asymmetry, and study its angular dependence.

We consider the external field specified by the vector potential~\eqref{eq:field} with the Gaussian envelope~\eqref{eq:F1}. We use the notation $\omega_1\equiv\Omega$ for the slow (strong) carrier frequency and $\omega_2\equiv\omega$ for the fast (weak) one. Unless stated otherwise, we focus on \emph{circular polarization} of both components in the $xy$
plane, i.e., $|\delta_1|=|\delta_2|=1$ in Eq.~\eqref{eq:field}. The baseline parameters are $E_1 = 0.2 E_{\text{c}}$, $E_2 = 0.04 E_{\text{c}}$, $\varphi_1=\varphi_2=0$, $\sigma=10$, and $\Omega=0.04m$, while the weak frequency $\omega$ is varied. When studying a single pulse, the amplitude of the other component is set to zero. The Keldysh parameters introduced in Sec.~\ref{sec:setup} are $\gamma_1 = 0.2$, $\gamma_2 = 25 (\omega/m)$, $\gamma_{\text{c}} = 5 (\omega/m)$.

\subsection{Spin-summed spectra} \label{subsec:ch2_spectra}

To characterize the spectra, we examine the spin-summed electron number density in momentum space~\eqref{eq:f_spin-summed}. We emphasize that the final kinetic momentum coincides with $\mathbf{p}$. In what follows, we visualize $f^{(e^-)}(\mathbf{p})$ in the $p_x p_y$ plane at $p_z=0$.

\subsubsection{Weak field only: multiphoton ring structure}
\label{subsubsec:ch2_weak}

\begin{figure}[t]
  \centering
  \includegraphics[width=0.99\linewidth]{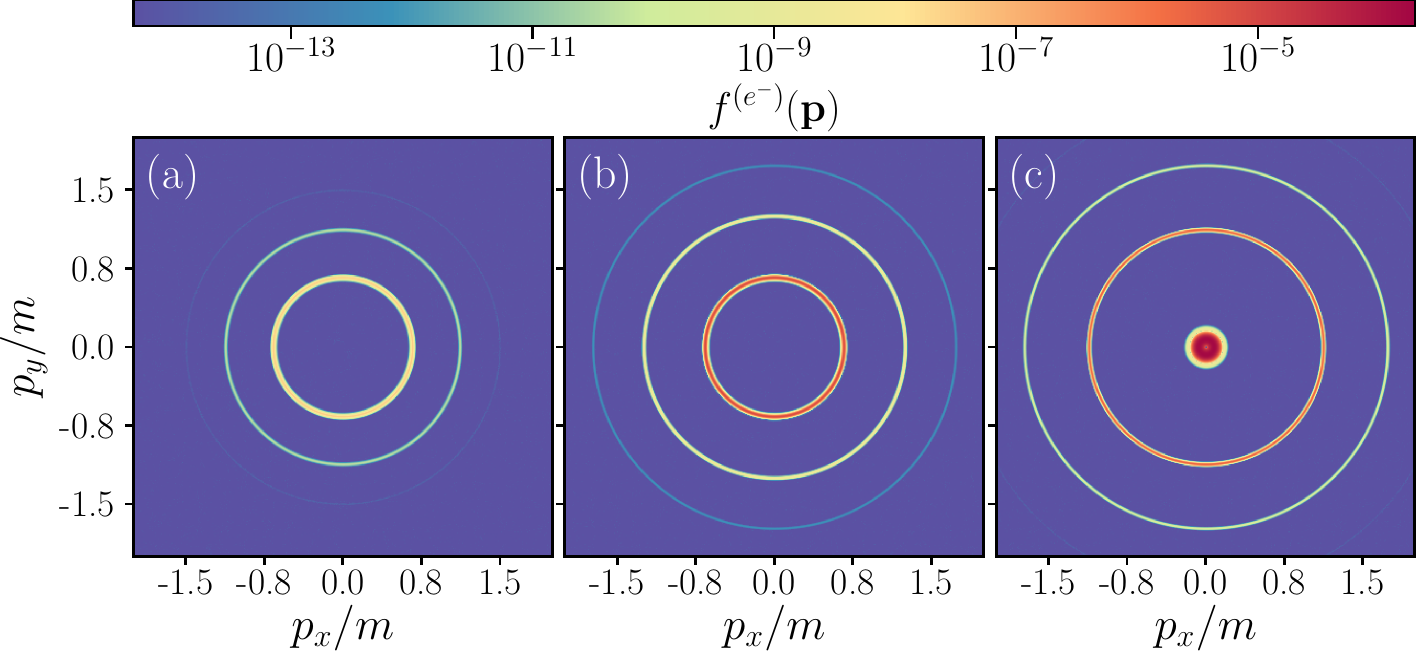}
  \caption{Spin-summed momentum distributions $f^{(e^-)}(\mathbf{p})$ produced by the \emph{weak field only} ($E_1=0$, $E_2=0.04E_{\text{c}}$) in the $p_x p_y$ plane at $p_z=0$. Panels: (a) $\omega=0.6m$, (b) $\omega=0.8m$, (c) $\omega=m$. The envelope parameter is $\sigma=10$ and the field is circularly polarized in the $xy$ plane. The ringlike maxima reflect multiphoton resonances characteristic of the perturbative regime ($\gamma_2\gg 1$).}
  \label{fig:weak_only}
\end{figure}

We first consider pair creation by the weak, rapidly oscillating pulse alone, i.e., we set $E_1=0$.
Since $\gamma_2\gg 1$ for the frequencies used below, the production mechanism is perturbative and the spectrum is expected to exhibit multiphoton resonances. Figure~\ref{fig:weak_only} shows the
spin-summed density $f^{(e^-)}(\mathbf{p})$ for three representative values of the weak frequency: $\omega=0.6m$, $0.8m$, and $m$.

The resulting spectra exhibit a pronounced ring structure, which can be explained in terms of multiphoton resonances. It is convenient to define a cycle-averaged quasienergy (effective energy)
associated with a given final momentum $\mathbf{p}$,
\begin{multline}
\mathcal{E}(\mathbf{p}) =
\frac{1}{2\pi}\int \limits_{0}^{2\pi}\! d\phi\,\Bigg[
m^2+p_z^2
+\Bigg(p_x+\frac{m}{\sqrt{2}\gamma_2}\cos\phi\Bigg)^{\!2}\\
+\Bigg(p_y+\frac{m}{\sqrt{2}\gamma_2}\sin\phi\Bigg)^{\!2}
\Bigg]^{1/2}.
\label{eq:effective_energy}
\end{multline}
The $\phi$-dependent terms reflect the ponderomotive, or effective-mass, dressing in the external field~\cite{aleksandrov_prd_2018,kohlfuerst_prl_2014}. We neglect the envelope in~\eqref{eq:effective_energy} because the pulse contains many carrier cycles for $\sigma=10$. In the perturbative regime $\gamma_2\gg 1$, the dressing is small and one may approximate
\begin{equation}
\mathcal{E}(\mathbf{p}) \approx \sqrt{m^2+\mathbf{p}^2}.
\label{eq:effective_energy_approx}
\end{equation}
The multiphoton resonance condition then reads
\begin{equation}
2 \mathcal{E}(\mathbf{p}_{\ngamma})=\ngamma\,\omega,
\label{eq:multi_resonance}
\end{equation}
yielding the ring radii
\begin{equation}
|\mathbf{p}_{\ngamma}|=\sqrt{\Big(\frac{\ngamma\,\omega}{2}\Big)^2-m^2}.
\label{eq:ring_radii}
\end{equation}
For $\omega=0.6m$, the rings in Fig.~\ref{fig:weak_only}(a) correspond to $\ngamma=4,5,6$, giving
$|\mathbf{p}_{4}|\approx 0.663m$, $|\mathbf{p}_{5}|\approx 1.12m$, and $|\mathbf{p}_{6}|\approx 1.50m$,
respectively, in excellent agreement with the numerical maxima (deviations below the percent level).
For $\omega=0.8m$ the dominant resonances are $\ngamma=3,4,5$, while for $\omega=m$ the visible rings correspond to $\ngamma=2,3,4,5$.

Thus, in the absence of the strong component, the momentum spectra in the perturbative regime exhibit a pronounced shell, or ring, structure that is well described by the multiphoton resonance condition~\eqref{eq:multi_resonance}. In the next subsection, we consider the spectra produced in the absence of the weak field.

\subsubsection{Strong field only: tunneling regime and turning-point mapping} \label{subsubsec:ch2_strong}

\begin{figure}[t]
  \centering
      \includegraphics[width=0.99\linewidth]{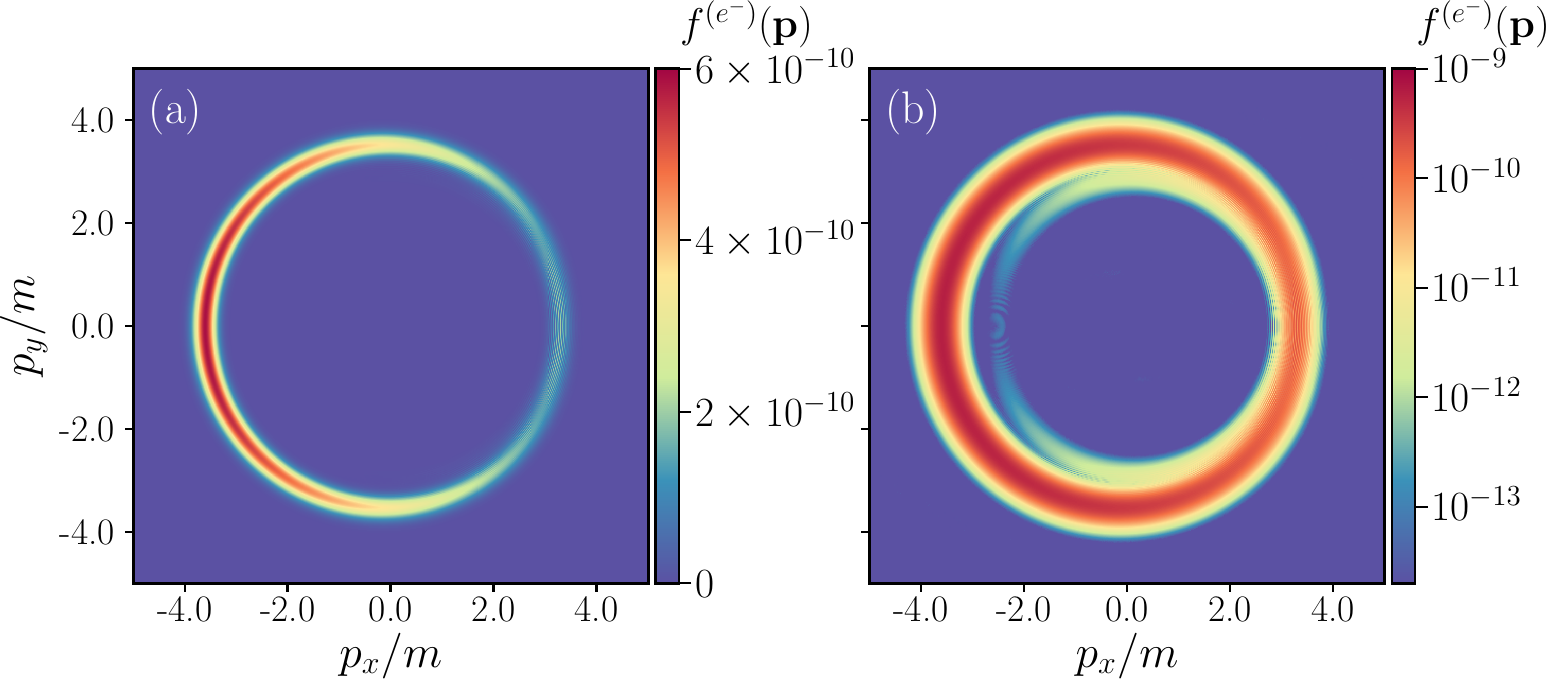}
  \caption{
  Spin-summed momentum distributions $f^{(e^-)}(\mathbf{p})$ of the electrons produced by the \emph{strong pulse only}
  ($E_1=0.2E_{\text{c}}$, $E_2=0$) in the $p_x p_y$ plane at $p_z=0$. Panel~(a): linear scale. Panel~(b): logarithmic scale highlighting the low-density tail. The remaining parameters are $\Omega=0.04m$ and $\sigma=10$.}
  \label{fig:strong_only}
\end{figure}

We next switch off the weak component ($E_2=0$) and consider the strong, slowly varying pulse alone with $E_1=0.2E_{\text{c}}$. The corresponding spectrum is shown in Fig.~\ref{fig:strong_only} in both linear and logarithmic scales. In contrast to the multiphoton case, the distribution is comparatively smooth, reflecting the tunneling nature of
pair creation for $\gamma_1\ll 1$.

A simple semiclassical picture explains the overall support and the brightest regions of the spectrum. Recall that the kinetic momentum evolves as $\mathbf{q}(t)=\mathbf{p}-e\mathbf{A}(t)$, and
$\mathbf{q}(t_{\text{out}})=\mathbf{p}$ in our gauge. In the tunneling regime, the
dominant contribution is expected to come from particles created with near-zero kinetic momentum at a ``creation time'' $t=t_{*}$. This leads directly to the mapping between the observed final momentum and the vector potential,
\begin{equation}
\mathbf{p} \approx e \mathbf{A}(t_{*}),
\label{eq:turning_point_map}
\end{equation}
so that the spectrum can be interpreted by scanning turning points $t_{*}$ over the pulse duration
and projecting them into momentum space. This procedure is described in detail, e.g., in  Refs.~\cite{aleksandrov_prd_2019_1,olugh_plb_2020,sevostyanov_prd_2021,aleksandrov_symmetry}.

The structure revealed in Fig.~\ref{fig:strong_only} is qualitatively reproduced within this semiclassical picture. This supports the standard interpretation that most particles are created with small kinetic momentum and are subsequently accelerated by the field, thereby acquiring the observed asymptotic momentum distribution (see, e.g., Refs.~\cite{kluger_prd_1992,blinne_gies_2014,aleksandrov_prd_2019_1,olugh_plb_2020,aleksandrov_kohlfuerst,sevostyanov_prd_2021,aleksandrov_symmetry}).

Moreover, if the external field varies slowly on the characteristic formation scale of the process, the turning-point picture can be turned into a quantitative estimate of the momentum distribution. This is achieved by employing the pair-production rate obtained for a constant background and evaluating it locally at the instantaneous field strength $\mathbf{E}(t_*)$. This procedure is commonly referred to as the locally constant field approximation (LCFA) and has been discussed extensively in the literature; see, for instance, Refs.~\cite{aleksandrov_prd_2019_1,aleksandrov_kohlfuerst,sevostyanov_prd_2021,aleksandrov_symmetry,aleksandrov_sevostyanov_2025,tkachev_pra_2025}. In the Appendix, we present several further observations concerning the LCFA in arbitrarily polarized fields. We stress already here that, although the LCFA can provide a very good description for sufficiently strong and slowly varying backgrounds, it necessarily misses interference effects and genuinely dynamical features of the pair-production process.

\subsubsection{Momentum distributions for the combined field} \label{subsec:combined_spectra}

We now consider the combined field with $E_1 = 0.2E_\text{c}$ and $E_2 = 0.04 E_\text{c}$. Since the total field strength is large, we expect the semiclassical picture outlined above to remain qualitatively valid. Once the weak component is added, the time dependence of $\mathbf{E}(t)$ becomes considerably richer, because the field acquires a large number of local extrema. Since the particle number density is governed by the local values of $\mathbf{E}(t_*)$ at the turning points $t_*$, the spectra should also exhibit additional oscillatory features as functions of~$\mathbf{p}$.

\begin{figure}[t]
  \centering
\includegraphics[width=\linewidth]{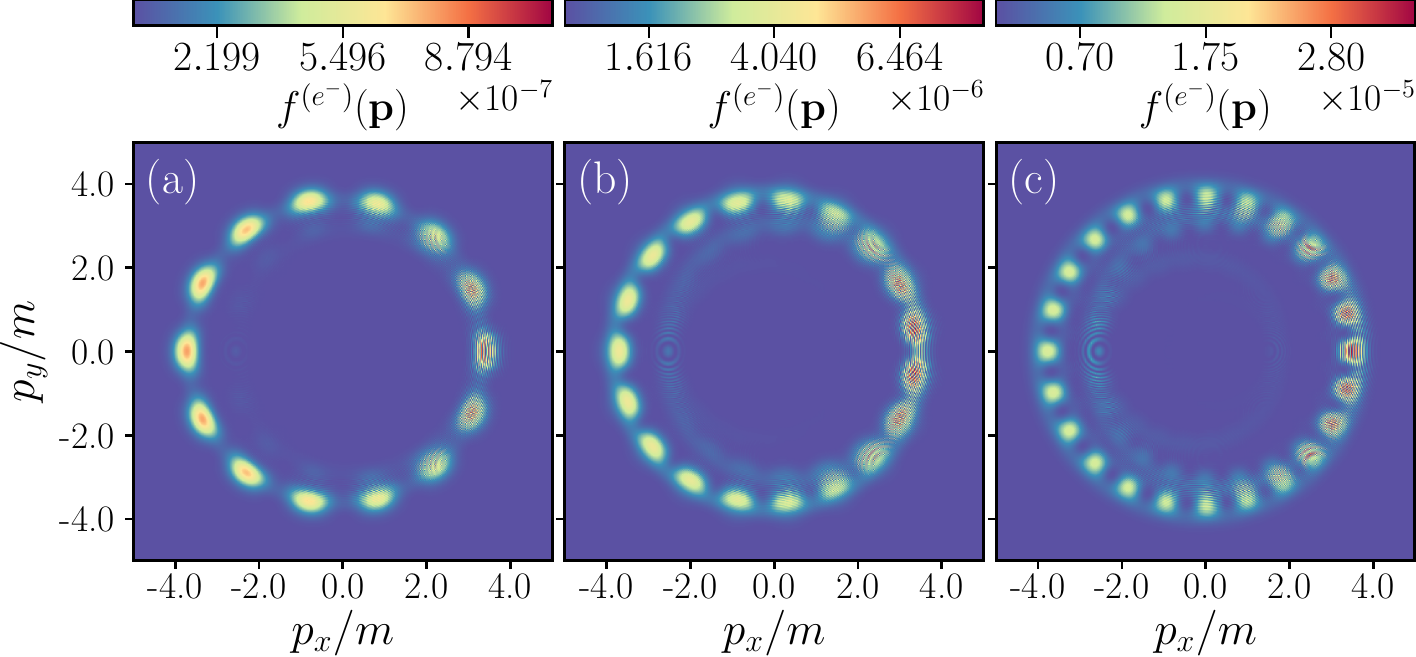}
  \caption{
  Spin-summed electron momentum distributions $f^{(e^-)}(\mathbf{p})$ in the \emph{combined} field
  \eqref{eq:field} with $E_1=0.2E_{\text{c}}$, $E_2=0.04E_{\text{c}}$, $\Omega=0.04m$ and $\sigma=10$, shown in the $p_x p_y$ plane at $p_z=0$.
  Panels: (a) $\omega=0.6m$, (b) $\omega=0.8m$, (c) $\omega=m$.
  The peak pattern follows local extrema of the time-dependent field strength, while the fine
  fringes reflect interference between contributions associated with multiple turning points.}
  \label{fig:combined_spectra}
\end{figure}

\begin{figure}[t]
  \centering
\includegraphics[width=\linewidth]{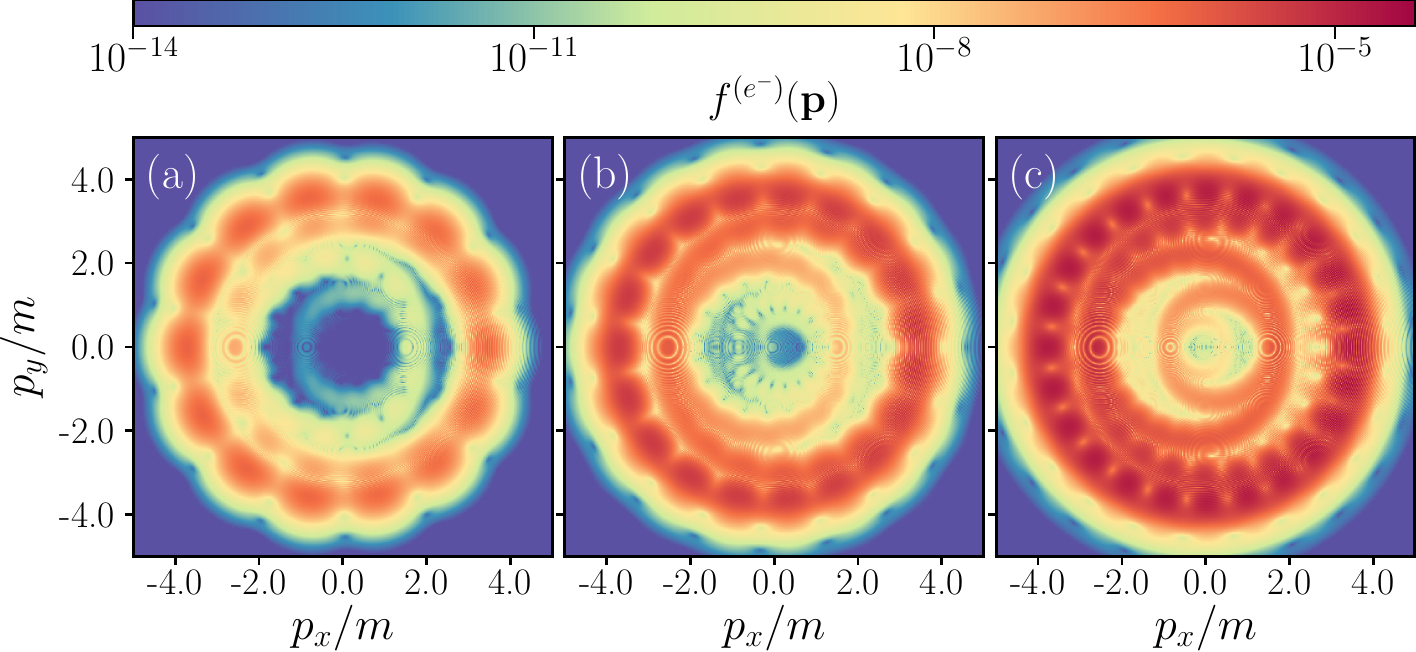}
  \caption{Logarithmic representation of the plots in Fig.~\ref{fig:combined_spectra}. The graphs resolve the internal structure of the inner rings and reveal fine interference fringes arising from the coexistence of multiple turning points contributing to the same final momentum $\mathbf{p}$.}
  \label{fig:combined_spectra_log}
\end{figure}

The corresponding electron momentum distributions obtained from the numerical solution of the QKE system are shown in Fig.~\ref{fig:combined_spectra}. The particle density in the combined field exceeds, by several orders of magnitude, the densities obtained for the strong and weak pulses separately, which clearly signals dynamical assistance. In addition, the plots display peaks associated with the local maxima of the field strength, in agreement with the semiclassical picture discussed above. To analyze the inner structures of the spectra, where the density is small, we also plot the distributions on a logarithmic scale (see Fig.~\ref{fig:combined_spectra_log}). Pronounced interference effects also become visible in the vicinity of the
intersections of the ringlike structures, where a given $\mathbf{p}$ is associated with several
turning points $t_*$~\cite{olugh_plb_2020}. As the weak-field frequency increases, the number of peaks in the electron spectra grows and the quantum-interference effects become significantly stronger. The spectra remain well resolved even in regions of very small densities, which indicates the numerical stability and accuracy of the computation. The qualitative structure of the momentum distributions can be captured within the LCFA, but in the presence of the weak fast pulse this approximation significantly underestimates the densities; see the Appendix for further details.

Having discussed the main properties of the spin-summed momentum distributions, we now turn to the helicity-resolved spectra, which constitute the central subject of the present study. Whereas Ref.~\cite{jiang_arxiv_2025} analyzes general spin projections in a single circularly polarized background, we focus here on helicity-resolved spectra in a dynamically assisted bichromatic rotating field and on the corresponding enhancement and angular organization of the helicity asymmetry.


\subsection{Helicity asymmetry of the momentum spectra}
\label{subsec:helicity_asym}

We now analyze the helicity-resolved electron spectra $f^{(e^-\mathrm{L/R})} (\mathbf{p})$ defined in Eq.~\eqref{eq:el_LR} and discuss the relation between the two helicity channels. In particular, we focus on the helicity asymmetry emerging in the momentum distributions. In what follows, we present the main results of our helicity-resolved analysis.

\begin{figure}[t]
  \centering
\includegraphics[width=\linewidth]{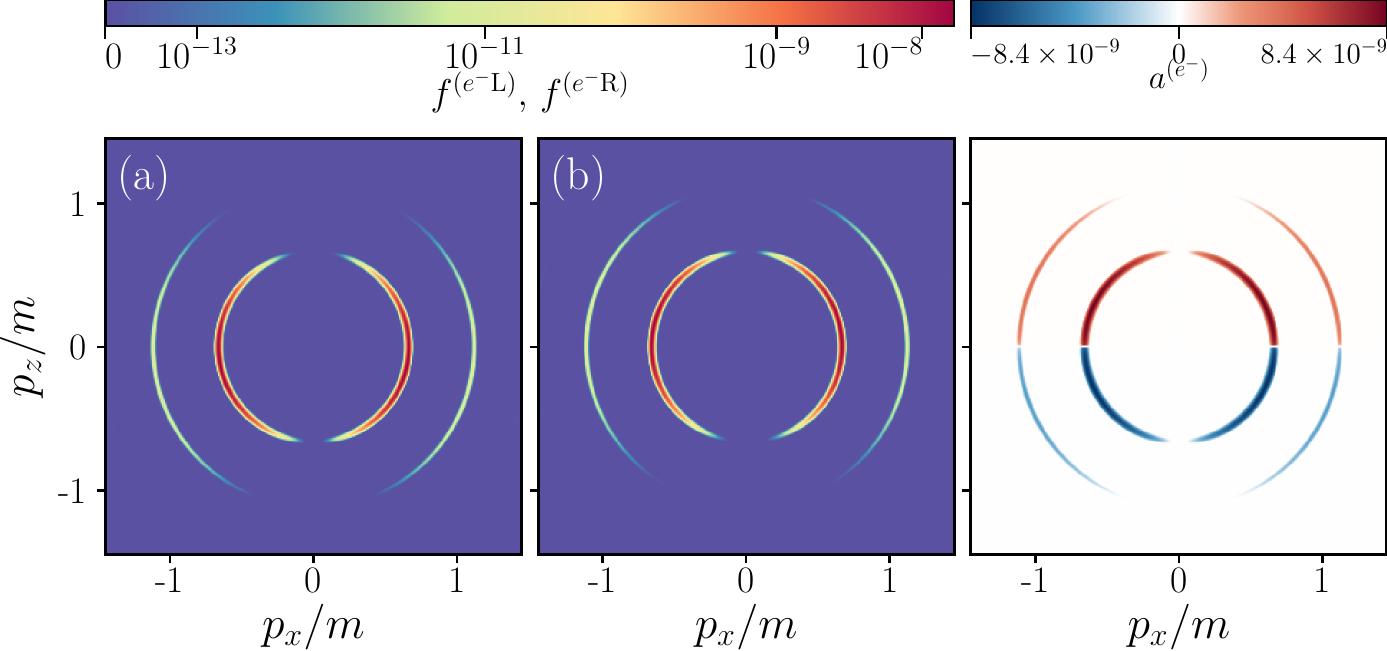}
\caption{
Helicity-resolved electron momentum distributions in the $p_x p_z$ plane at $p_y=0$ for the
\emph{weak field only} configuration ($E_1=0$).
Panels: (a)~negative helicity, $f^{(e^-\mathrm{L})}$; (b)~positive helicity, $f^{(e^-\mathrm{R})}$;
(c)~difference $f^{(e^-\mathrm{R})}-f^{(e^-\mathrm{L})}$.}
  \label{fig:LR_xz_weak}
\end{figure}

\subsubsection{Fixed-helicity momentum spectra}
\label{subsubsec:fixed_helicity_spectra}

Throughout this subsection, we fix the frequency of the fast (weak) component to $\omega=0.6m$ and present helicity-resolved spectra for the weak field only, the strong field only, and the combined field. In the present setup, helicity effects are most clearly visible in momentum planes with $p_z\neq 0$. We therefore discuss the asymptotic distributions in the $p_x p_z$ plane. In what follows we present results for circular polarization,
$\delta_1=\delta_2=1$ (the linearly polarized case is less informative here because of its axial symmetry with respect to the $x$ axis).

Figure~\ref{fig:LR_xz_weak} shows the helicity-resolved spectra of electrons produced by the \emph{weak}
field alone, together with the difference between the right- and left-handed distributions.
All spectra in Fig.~\ref{fig:LR_xz_weak} are computed at $p_y=0$. As expected from the symmetry properties discussed above, the difference $a^{(e^-)} (\mathbf{p}) \equiv f^{(e^-\mathrm{R})} (\mathbf{p})-f^{(e^-\mathrm{L})} (\mathbf{p})$ is an odd function of the momentum projection
$p_z$, in agreement with the symmetry properties discussed previously and observed in
Ref.~\cite{aleksandrov_kudlis_prdl_2024}. In particular, right-handed electrons predominantly acquire a
positive $p_z$ component, while left-handed electrons are mainly produced with negative $p_z$.

In this weak-field case, the spectral structure is still fully described by the photon-resonance
picture discussed earlier: the spectra exhibit pronounced ``shells'' corresponding to multiphoton
channels. For $\omega=0.6m$ the most clearly visible shells correspond to $\ngamma=4$ and $\ngamma=5$.
At the same time, this resonance-based picture does not by itself explain the suppression of the distributions with increasing $|p_z|$.

\begin{figure}[t]
  \centering \includegraphics[width=\linewidth]{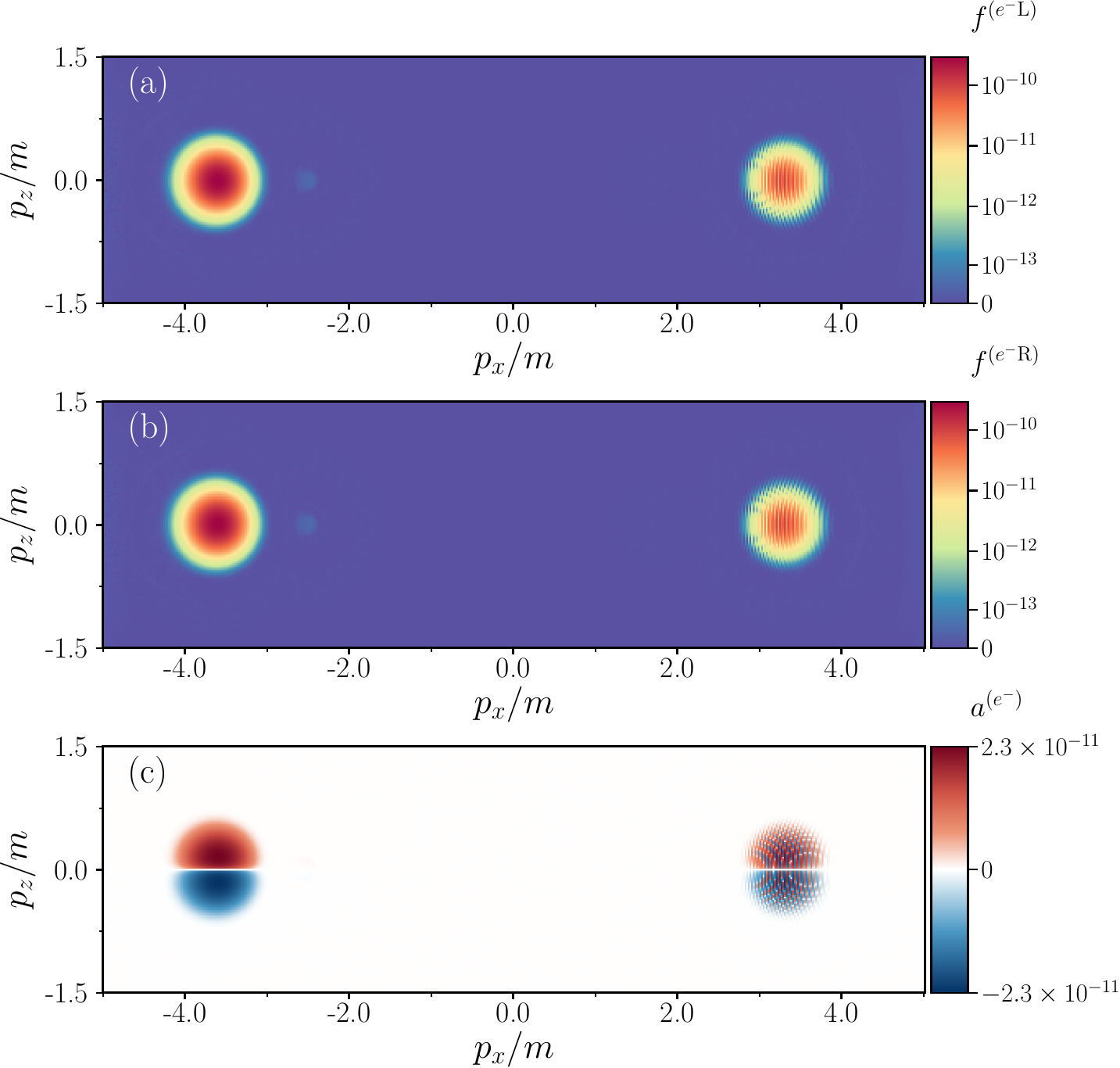}
  \caption{Helicity-resolved electron momentum distributions in the $p_x p_z$ plane for the \emph{strong
  field only} ($E_2=0$). Panels: (a)~negative helicity (L); (b)~positive helicity (R); (c)~difference $f^{(e^-\mathrm{R})}-f^{(e^-\mathrm{L})}$. The distributions are shown using the nonlinear scale~\eqref{eq:ASH}. The arcsinh-based representation resolves the fine low-density structure while remaining close to a logarithmic scaling in the relevant data range.}
  \label{fig:LR_xz_strong}
\end{figure}

\begin{figure}[t]
  \centering
\includegraphics[width=\linewidth]{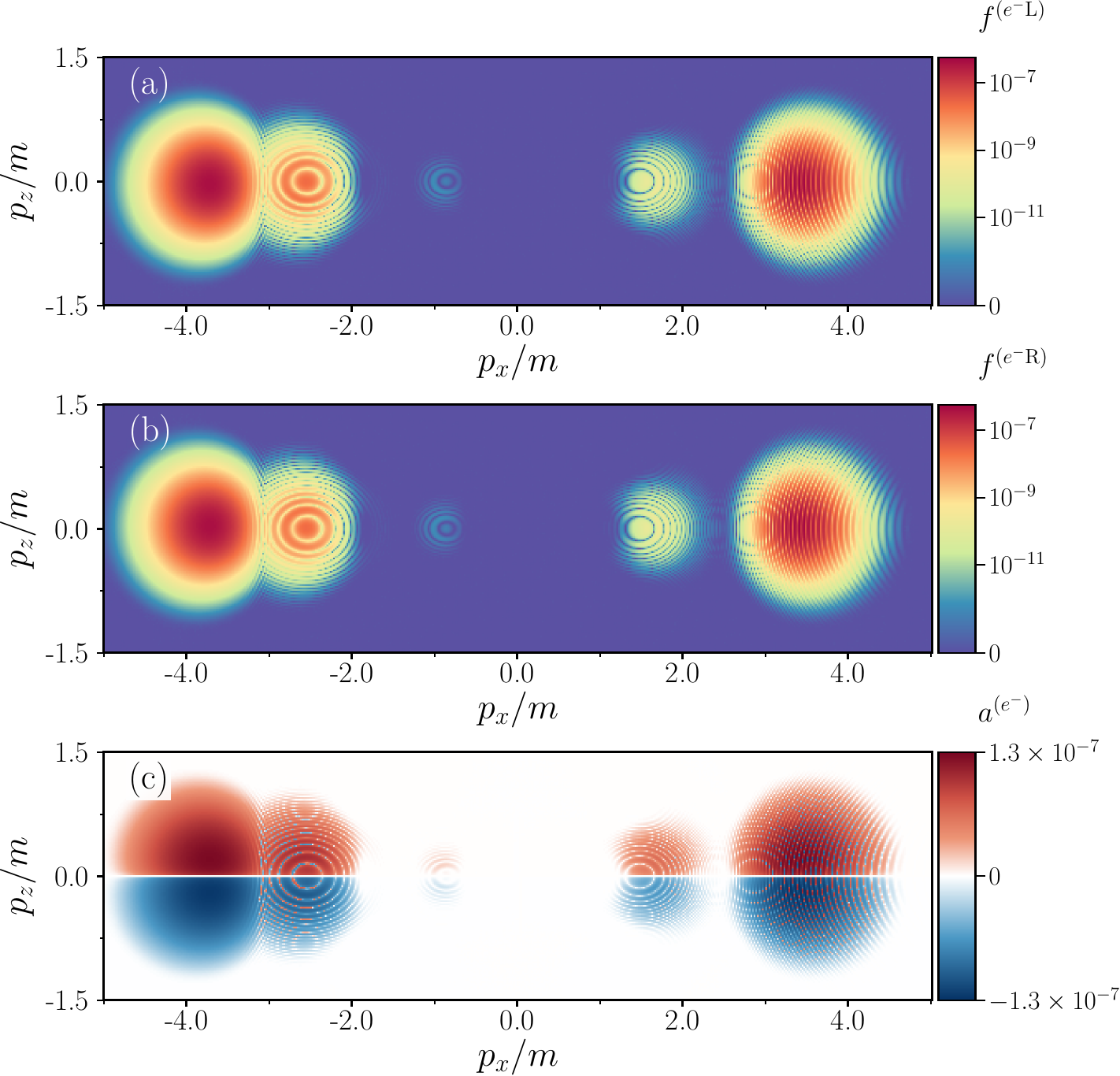}
  \caption{Helicity-resolved electron momentum distributions in the $p_x p_z$ plane for the
  \emph{combined} strong and weak fields. Panels: (a)~negative helicity (L); (b)~positive helicity (R); (c)~difference $f^{(e^-\mathrm{R})}-f^{(e^-\mathrm{L})}$. The distributions are shown using the nonlinear scale~\eqref{eq:ASH}. The addition of the fast weak component modifies the spectral pattern and enhances the helicity imbalance compared to the strong-field case.}
  \label{fig:LR_xz_comb}
\end{figure}

The corresponding helicity-resolved results for the \emph{strong field only} are shown in Fig.~\ref{fig:LR_xz_strong},
and the corresponding spectra for the \emph{combined} fields are presented in
Fig.~\ref{fig:LR_xz_comb}. For the strong field and for the combined field, the characteristic suppression
of the spectra with increasing $|p_z|$ can be understood semiclassically: since $p_z$ is orthogonal to the electric-field plane and contributes to the effective transverse energy scale, larger $|p_z|$ corresponds to a larger effective energy cost and hence to a smaller production probability. To visualize both the high- and low-density regions of $a^{(e^-)}(\mathbf{p}) = f^{(e^-\mathrm{R})}(\mathbf{p})-f^{(e^-\mathrm{L})}(\mathbf{p})$, we use a nonlinear scale based on the inverse hyperbolic sine,
\begin{equation}
y(x)=\sinh^{-1}(A x),
\label{eq:ASH}
\end{equation}
with $A$ chosen such that, within our data range, the scale is close to a logarithmic one while remaining well behaved near zero.

\begin{figure}[t]
  \centering
    \includegraphics[width=\columnwidth]{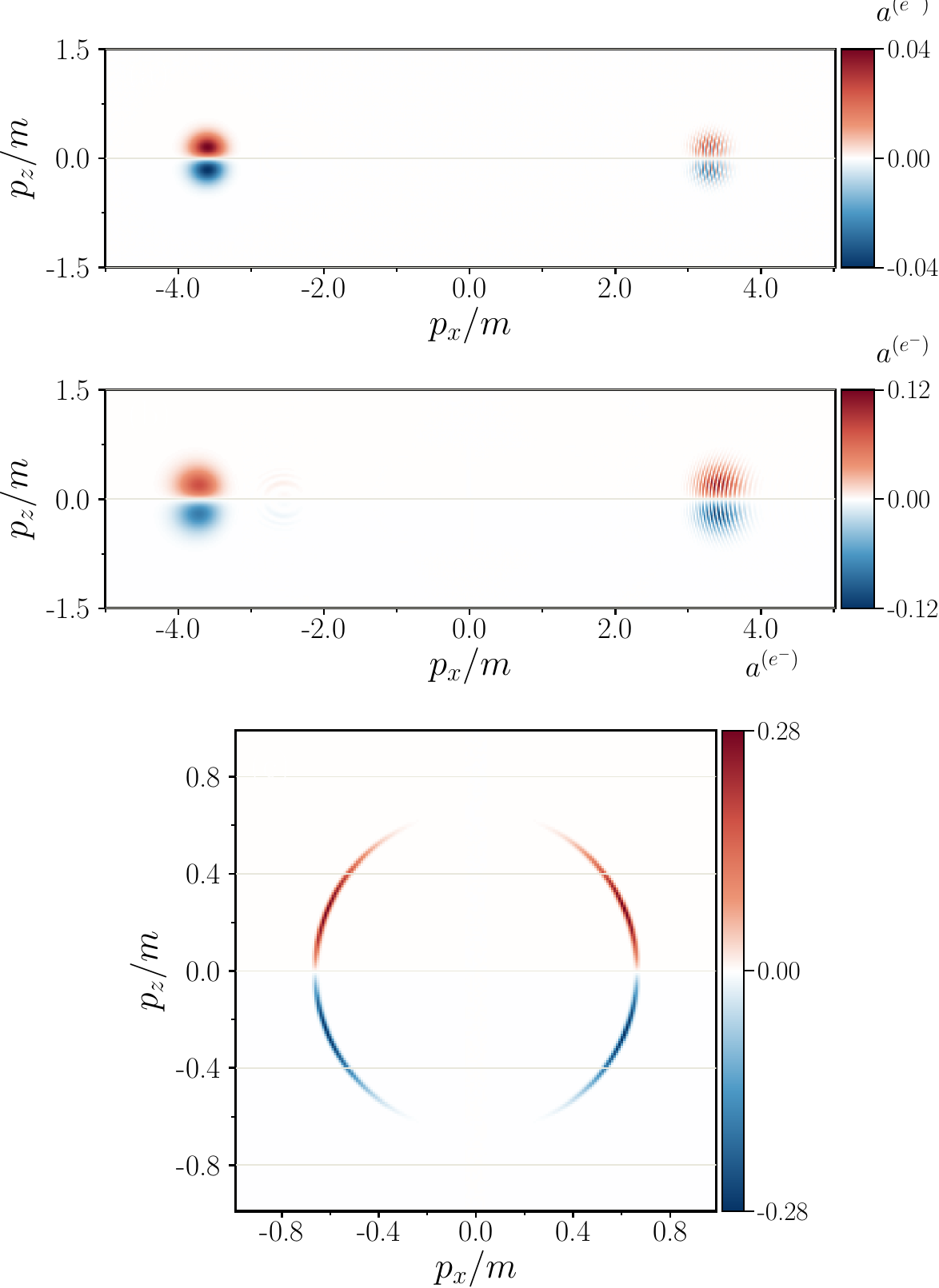}
\caption{Helicity-asymmetry degree $a^{(e^-)} = f^{(e^-\mathrm{R})}-f^{(e^-\mathrm{L})}$ in the $p_x p_z$ plane at $p_y=0$ ($\omega=0.6m$). The function $a^{(e^-)}$ is normalized here by the maximal
spin-summed electron density within the same plane (the normalization is performed separately in each panel).
Panels: (a) strong field only ($E_2=0$); (b) combined field ($E_1=0.2E_{\text{c}}$, $E_2=0.04E_{\text{c}}$);
(c) weak fast-oscillating field only ($E_1=0$).}
  \label{fig:a_summary}
\end{figure}

To quantify the helicity asymmetry, Fig.~\ref{fig:a_summary} shows the asymmetry degree $a^{(e^-)}$, defined as the difference $f^{(e^-\mathrm{R})}-f^{(e^-\mathrm{L})}$ normalized by the
maximum value of the (spin-summed) electron density in the corresponding momentum plane.
We observe that, while in the strong-field case the asymmetry remains below about $4\%$, the addition of the weak fast-oscillating component enhances the asymmetry up to about $10\%$. We also note that the largest asymmetry values, reaching about $25\%$--$30\%$, occur in the multiphoton regime. This behavior is related to the shell structure of the spectrum, which allows one to
expect production of electrons with comparatively large values of $|p_z|$. Since the difference
between the right- and left-handed densities increases rapidly with $|p_z|$, the weak-field
(multiphoton) scenario can exhibit a strong helicity asymmetry. Increasing the frequency of the field in this regime is found to further enhance the observed effect.

In addition to helicity-resolved spectra, one may also consider spin asymmetries defined with respect to a fixed laboratory axis. Such observables can exhibit comparatively large asymmetry degrees in certain field configurations; see, e.g., Ref.~\cite{hu_prd_2024}. More generally, fixed-axis spin projections and helicity correspond to different special cases of spin-resolved momentum distributions~\cite{jiang_arxiv_2025}. In the present setup, however, helicity is the more natural observable, since it is tied to the particle propagation direction and hence to the preferred axis selected by the rotating background.

\subsubsection{Azimuthal distributions}
\label{subsubsec:angular_distributions}

For the analysis of helicity effects it is convenient to use spherical coordinates $(|\mathbf{p}|,\theta,\varphi)$ for the asymptotic kinetic momentum $\mathbf{p}$. Here $\theta$ is the polar angle measured from the $z$ axis, while $\varphi$ is the azimuthal angle in the $xy$ plane. Since the external electric field rotates in the $xy$ plane, the $z$ axis plays the role of the distinguished direction, and it is therefore natural to analyze the helicity-resolved spectra in terms of their angular dependence. To this end, we integrate over the momentum magnitude $|\mathbf{p}|$ and define angular densities as functions of $\theta$ and $\varphi$:
\begin{equation}
f^{(e^-\text{L/R})}_{\theta,\varphi}(\theta,\varphi)=\frac{1}{m^3}\int\limits_0^{\infty} d|\mathbf{p}|\,|\mathbf{p}|^2 f^{(e^-\text{L/R})}(\mathbf{p}).
\label{eq:f_theta_phi}
\end{equation}
These quantities allow us to separate the overall spectral weight from the angular structure of the helicity-resolved distributions.

\begin{figure}[t]
  \centering
\includegraphics[width=\linewidth]{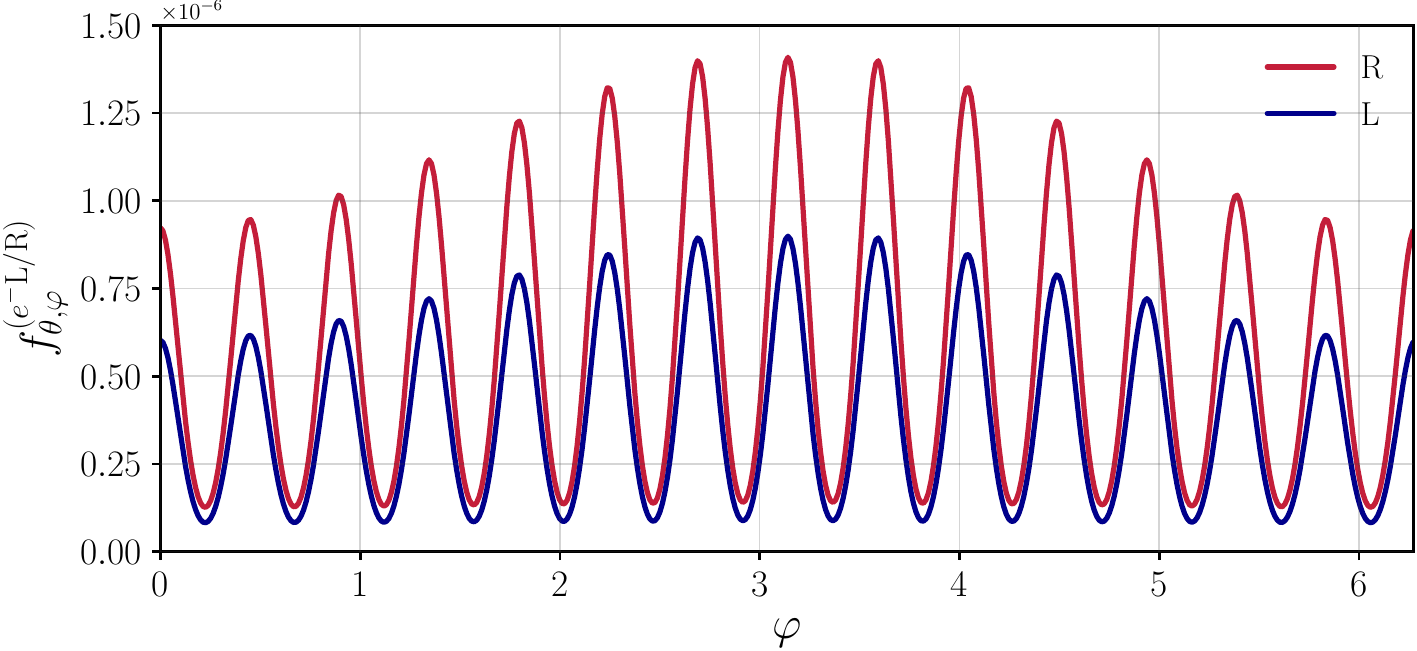}
\caption{Azimuthal distribution of electrons as a function of $\varphi$ at fixed polar angle
$\theta=86^\circ$ for the combined field~\eqref{eq:field} with $\omega=0.6m$.
The two curves correspond to positive (R) and negative (L) helicity states. Their nearly identical shape, up to an overall factor, shows that the helicity ratio depends only weakly on the azimuthal angle at fixed $\theta$.}
\label{fig:LR_phi}
\end{figure}

Figure~\ref{fig:LR_phi} displays the dependence of the angular density on the azimuthal angle $\varphi$ at $\theta=86^\circ$ for the combined field and for both helicities. A key observation is that the ratio of the densities for positive and negative helicity is largely insensitive to $\varphi$: one curve is reproduced from the other by multiplication with an almost constant factor. This shows that the two helicity-resolved distributions are very similar in their azimuthal shape and differ primarily by an overall normalization.

A further important point is that the same ratio is also found to depend only weakly on the momentum magnitude $|\mathbf{p}|$; see also Fig.~\ref{fig:LR_momentum} and the discussion below. Therefore, the proportionality coefficient $P$ need not be extracted from the integrated angular densities~\eqref{eq:f_theta_phi}. Instead, it can be determined directly from the ratio of the helicity-resolved distributions at fixed momentum $\mathbf{p}$. In practice, it is advantageous to choose $\mathbf{p}$ in the vicinity of spectral maxima in order to improve numerical stability and accuracy.

Thus, despite the strongly nonlinear character of the pair-production process, the ratio of the densities for opposite helicities is not governed by a complicated dependence on all momentum variables. Rather, its structure is controlled predominantly by the polar angle $\theta$, while the dependences on $\varphi$ and $|\mathbf{p}|$ are comparatively weak. This observation considerably simplifies the description of the helicity asymmetry and naturally motivates a more detailed analysis of its polar-angle dependence, to which we now turn.

\subsubsection{Dependence of helicity asymmetry on the polar angle $\theta$}
\label{subsubsec:theta_dependence}

\begin{figure}[t]
  \centering
 \includegraphics[width=0.99\linewidth]{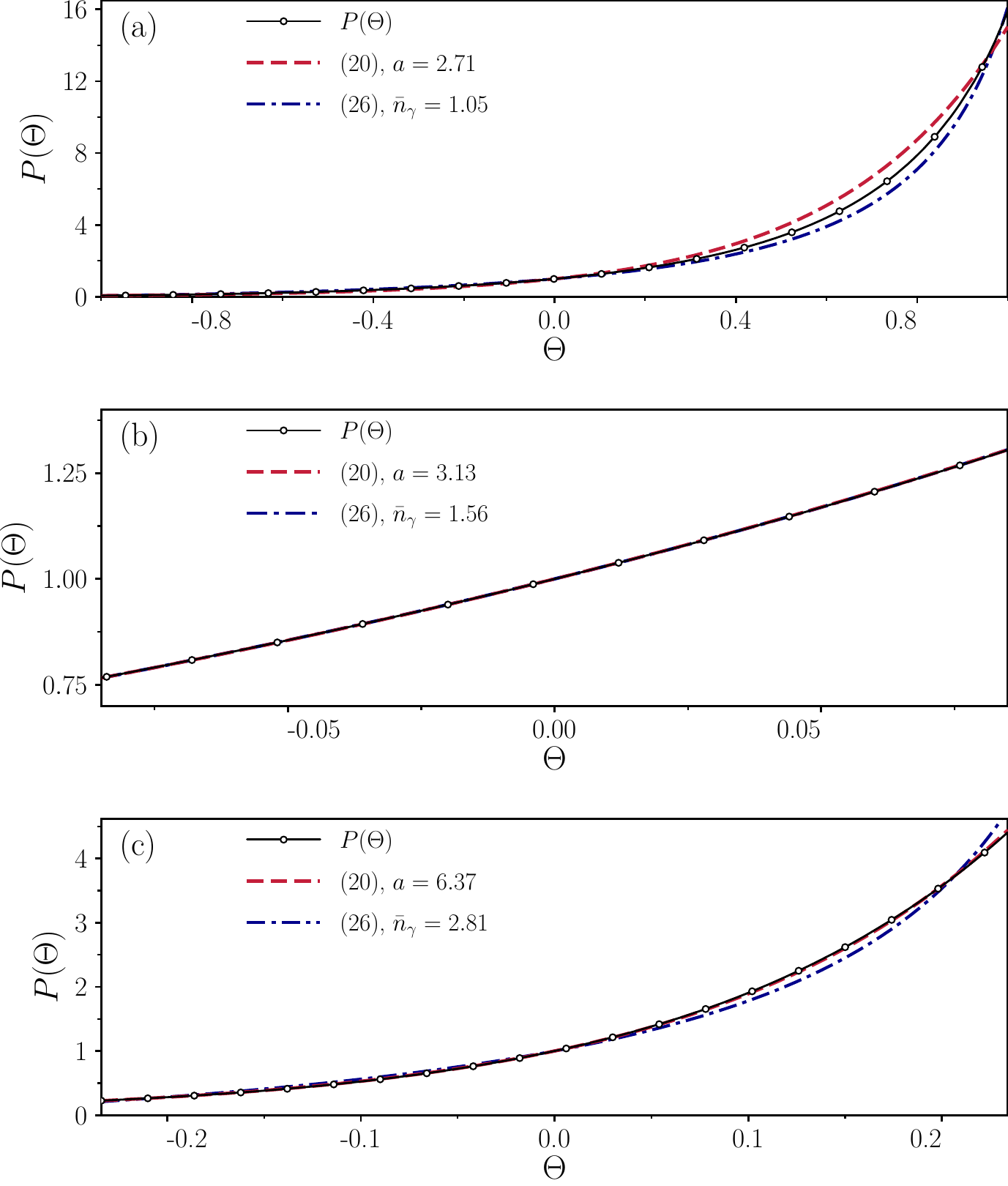}
\caption{Ratio of the helicity-resolved spectra, $P(\Theta)=f^{(e^-\mathrm{R})}/f^{(e^-\mathrm{L})}$ as a function of
$\Theta=\pi/2-\theta$ for $\omega=0.6m$.
Panels: (a) weak field only ($E_1=0$); (b) strong field only ($E_2=0$); (c) combined field. Solid curves show the QKE results. Dashed curves correspond to an exponential fit of the form $P(\Theta)=\mathrm{exp}(a\Theta)$ with fit parameter $a$. Dash-dotted curves represent the qualitative expectation based on Eq.~\eqref{eq:exp_idea}. The angular ranges differ between panels due to the different
spectral decay with $|p_z|$ and the resulting numerical sensitivity of the ratio.}
  \label{fig:P_a_n}
\end{figure}

We now investigate how the proportionality coefficient $P$ depends on $\theta$ at fixed field parameters. In Fig.~\ref{fig:P_a_n} we present the ratio of the spectra for positive and negative helicity as a function of the angle $\Theta\equiv \pi/2-\theta$, which measures the deviation of $\theta$ from $90^\circ$, for the weak field, the strong field, and the combined field, in the same order as the panels. For the weak-field case we show the results within a wider angular interval. This is because the weak-field spectrum decays
more slowly with increasing $|p_z|$, which substantially simplifies the numerical evaluation of the ratio at larger $|\Theta|$. Crucially, we observe that adding the weak field not only increases the total yield through dynamical assistance, but also significantly enhances the \emph{relative} helicity asymmetry. By fitting the ratios computed for a set of angles $\Theta$, we find that within the shown range the dependence can be well described by
\begin{equation}
P(\Theta)=\mathrm{e}^{a \Theta},
\label{eq:P_theta}
\end{equation}
where the parameter $a=a(E_1,E_2,\Omega,\omega,\sigma)$ is determined by the specific field configuration and provides a compact quantitative measure of the helicity asymmetry. In particular, a larger value of $a$ corresponds to a stronger separation of the two helicity channels as $\Theta$ increases. In the combined-field case this parameter is larger than in the strong-field-only case, showing that dynamical assistance enhances not only the total yield but also the angular helicity selectivity. We also note that the property $P(-\Theta) = 1/P(\Theta)$ matches the $p_z$-reflection behavior revealed previously, e.g., in the $p_x p_z$ momentum distributions (it is an exact symmetry of the QKE system in our external-field setup). This property implies that the helicity-asymmetry ratio must satisfy
\begin{equation}
P(\Theta) = \frac{1+g(\Theta)}{1-g(\Theta)}
\label{eq:P_g}
\end{equation}
for some odd function $g(\Theta)$. Our heuristic choice~\eqref{eq:P_theta} corresponds to $g(\Theta) = \tanh (a\Theta/2)$.

To obtain a qualitative understanding of the $P(\Theta)$ pattern, we now employ a simple angular-momentum argument. The discussion below is intended as a heuristic interpretation of the observed trend rather than a derivation of the exact fitted dependence. Consider the angular-momentum conservation law for a pair created in a circularly polarized field,
\begin{multline}
\mathbf{L}^{(e^-)}(\mathbf{p})+\mathbf{S}^{(e^-)}(\mathbf{p})\\
+\mathbf{L}^{(e^+)}(-\mathbf{p})
+\mathbf{S}^{(e^+)}(-\mathbf{p})
= \ngamma \mathbf{e}_z,
\label{eq:LMKD}
\end{multline}
where $\mathbf{L}$ and $\mathbf{S}$ are the orbital and spin angular momenta of the particles,
respectively. The left-hand side is the total angular momentum of the produced electron--positron
pair, whereas the right-hand side corresponds to the angular momentum of $\ngamma$ absorbed quanta of the
circularly polarized external field (the $z$ projection of each quantum is $+1$). Taking the scalar product of Eq.~\eqref{eq:LMKD} with the unit momentum vector
$\mathbf{p}/|\mathbf{p}|$ and using $(\mathbf{L},\mathbf{p})=0$, we obtain
\begin{equation}
\frac{h^{(e^-)}}{2}-\frac{h^{(e^+)}}{2}=\ngamma\cos\theta,
\label{eq:cos}
\end{equation}
where $h=\pm 1$ is the helicity of the electron or positron.

A corresponding relation may be written at the level of ensemble-averaged quantities. Therefore one can write
\begin{multline}
\frac{1}{2}\Bigg[
\frac{f^{(e^-\mathrm{R})}(\mathbf{p})-f^{(e^-\mathrm{L})}(\mathbf{p})}
     {f^{(e^-\mathrm{R})}(\mathbf{p})+f^{(e^-\mathrm{L})}(\mathbf{p})}\\-
\frac{f^{(e^+\mathrm{R})}(-\mathbf{p})-f^{(e^+\mathrm{L})}(-\mathbf{p})}
     {f^{(e^+\mathrm{R})}(-\mathbf{p})+f^{(e^+\mathrm{L})}(-\mathbf{p})}
\Bigg]
= \nbgamma\cos\theta,
\label{eq:medium}
\end{multline}
where $\nbgamma$ is the mean number of absorbed photons responsible for pair creation at the considered
time. Using the positron--electron relation \eqref{eq:pos_LR}, we arrive at
\begin{equation}
\frac{f^{(e^-\mathrm{R})}(\mathbf{q})-f^{(e^-\mathrm{L})}(\mathbf{q})}
     {f^{(e^-\mathrm{R})}(\mathbf{q})+f^{(e^-\mathrm{L})}(\mathbf{q})}
= \nbgamma\cos\theta.
\label{eq:medium_electrons}
\end{equation}
Introducing now
$P(\mathbf{p})\equiv f^{(e^-\mathrm{R})}(\mathbf{p})/f^{(e^-\mathrm{L})}(\mathbf{p})$
and switching to $\Theta=\pi/2-\theta$, we obtain
\begin{equation}
P(\mathbf{p})=P(\Theta)=\frac{1+\nbgamma\sin\Theta}{1-\nbgamma\sin\Theta}.
\label{eq:exp_idea}
\end{equation}
Equation~\eqref{eq:exp_idea} suggests that, within this heuristic picture, the ratio of the momentum spectra for opposite helicities is largely insensitive to the momentum magnitude and to the azimuthal angle $\varphi$. This provides a qualitative explanation for the observed similarity of the distributions. At the same time, the functional form~\eqref{eq:exp_idea} satisfies Eq.~\eqref{eq:P_g} but differs from the purely exponential fit~\eqref{eq:P_theta} obtained from the numerical data (see Fig.~\ref{fig:P_a_n}).

\begin{figure}[t]
  \centering
  \includegraphics[width=0.99\linewidth]{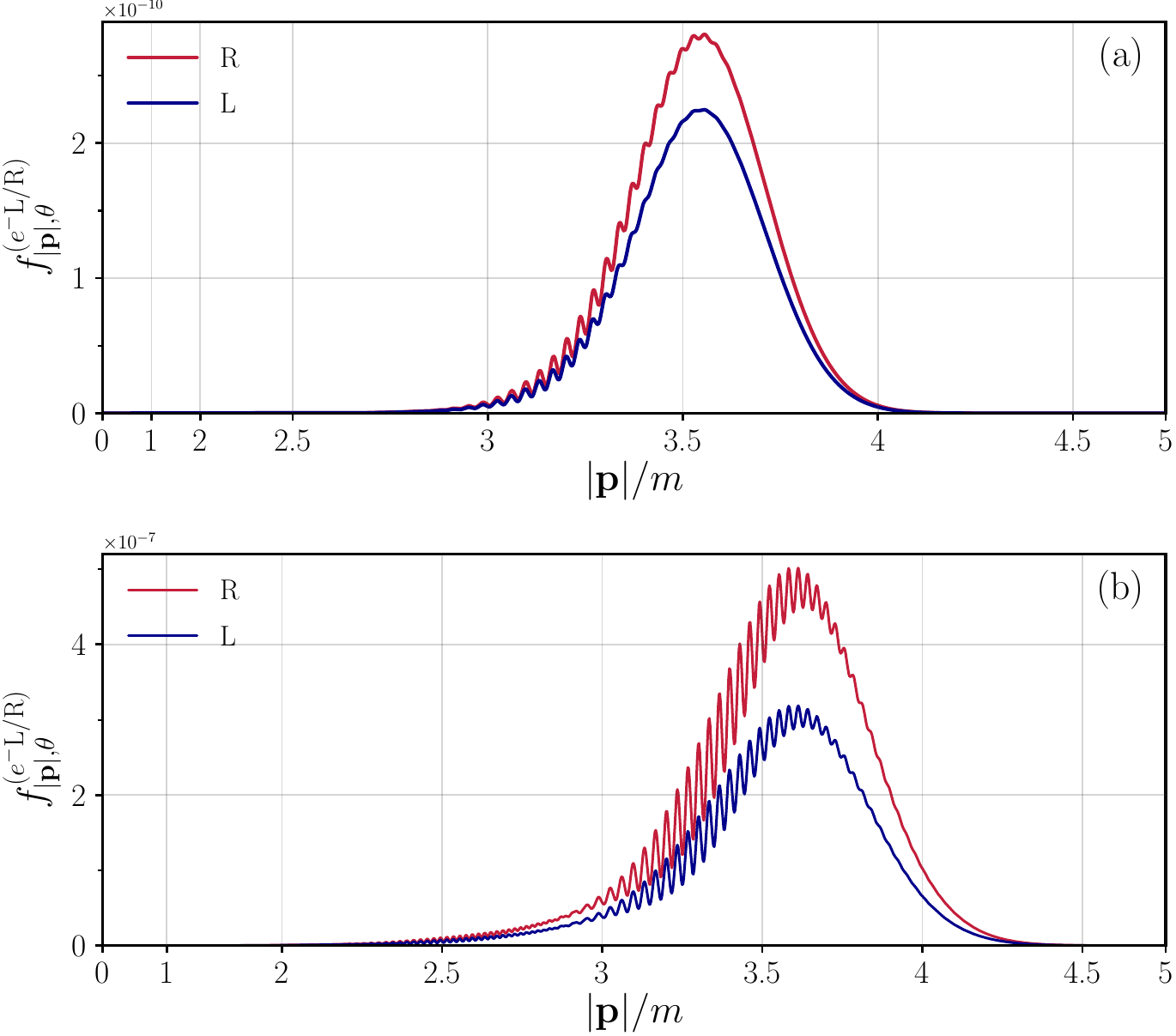}
  \caption{Helicity-resolved radial profiles $f^{(e^-\text{L/R})}_{|\mathbf{p}|, \theta}$ at fixed polar angle
  $\theta=86^\circ$. Panel~(a): strong field only
  ($E_1=0.2E_{\mathrm c}$, $E_2=0$). Panel~(b): combined field
  ($E_1=0.2E_{\mathrm c}$, $E_2=0.04E_{\mathrm c}$, $\omega=0.6m$). The red and blue curves denote
  positive (R) and negative (L) electron helicities, respectively. In each panel the two helicity
  channels have the same radial structure, showing that the helicity ratio is governed mainly by $\theta$ and depends only weakly on $|\mathbf{p}|$.}
  \label{fig:LR_momentum}
\end{figure}

For small $|\Theta|$, Eq.~\eqref{eq:exp_idea} is equivalent to $\mathrm{exp}(2\nbgamma\Theta)$, which indicates that fitting the data within a very narrow $\Theta$ window yields $a=2\nbgamma$. In this limit, the two fit curves become practically indistinguishable, as is indeed observed in Fig.~\ref{fig:P_a_n}(b). In relatively large intervals, Eqs.~\eqref{eq:P_theta} and \eqref{eq:exp_idea} are quantitatively different.

To confirm that the helicity-asymmetry ratio $P$ depends only weakly on the radial momentum coordinate, we also compute the $\varphi$-integrated distributions,
\begin{equation}
f^{(e^-\text{L/R})}_{|\mathbf{p}|, \theta} (|\mathbf{p}|, \theta) = \int\limits_0^{2\pi} d\varphi \, f^{(e^-\text{L/R})} (\mathbf{p}).
\end{equation}
The radial-momentum distributions for $\theta=86^\circ$ are presented in Fig.~\ref{fig:LR_momentum}. For both the strong-field-only and the combined-field configurations, the positive- and negative-helicity profiles have the same peak positions and oscillatory structure as functions of $|\mathbf{p}|$; the
helicity change mainly rescales the common radial envelope. The combined field also raises the absolute density by several orders of magnitude in the plotted range, consistently with dynamical assistance. The $|\mathbf{p}|$ distributions are consistent with the conclusion that dynamical assistance enhances the relative helicity imbalance as well.

\begin{figure}[t]
  \centering
  \includegraphics[width=0.99\linewidth]{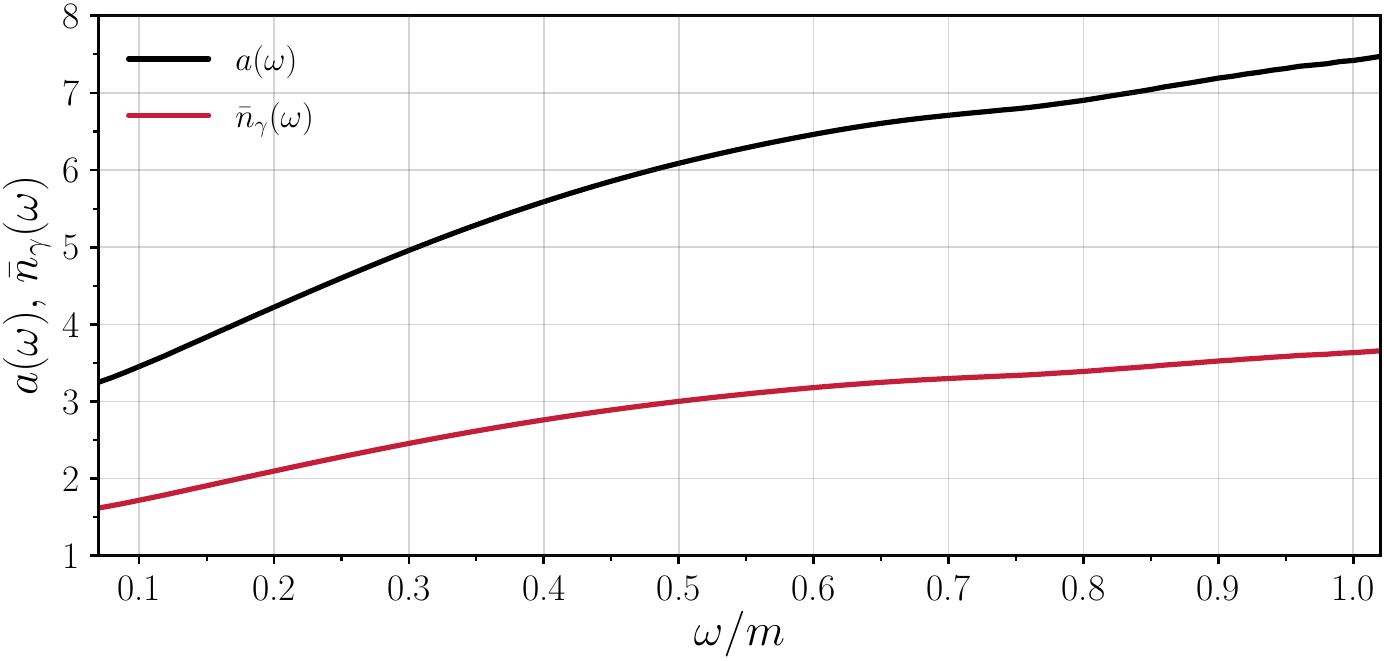}
  \caption{Frequency dependence of the fitted angular-asymmetry exponent $a(\omega)$ from
  $P(\Theta)=\exp[a(\omega)\Theta]$ and of the effective parameter $\nbgamma(\omega)$ entering Eq.~\eqref{eq:exp_idea}, evaluated for the baseline
  combined-field setup while varying the weak-field frequency $\omega$. The approximate relation
  $a(\omega)\simeq 2\nbgamma(\omega)$ follows from the small-$|\Theta|$ expansion of Eq.~\eqref{eq:exp_idea}.}
  \label{fig:a_n_omega}
\end{figure}

The frequency dependence of the fit parameters extracted in a small $\Theta$ window is shown in Fig.~\ref{fig:a_n_omega}. For the baseline combined-field setup, both the fitted exponent $a(\omega)$ in Eq.~\eqref{eq:P_theta} and the effective angular-momentum parameter $\nbgamma(\omega)$ entering Eq.~\eqref{eq:exp_idea} increase with the weak-field frequency. The plotted curves also show that $a(\omega)$ remains close to $2\nbgamma(\omega)$, as expected from the small-angle expansion of Eq.~\eqref{eq:exp_idea}. This provides a compact quantitative characterization of how the fast mode strengthens the helicity selectivity of the produced pairs.

In summary, despite the strong nonlinearity of the pair-production process in the considered field, the helicity-resolved spectra are found to be approximately proportional to each other. Thus, the helicity imbalance in the assisted rotating field is governed predominantly by the polar angle, with only weak residual dependence on $|\mathbf{p}|$ and $\varphi$, which provides a simple organizing principle for the helicity-resolved spectra.


\section{Conclusions}
\label{sec:concl}

We investigated vacuum $e^-e^+$ pair production in a spatially uniform bichromatic electric field within the corrected quantum-kinetic framework for fermions~\cite{aleksandrov_kudlis_klochai,aleksandrov_arxiv_2026}. The considered background models, in the dipole approximation, the superposition of two counterpropagating circularly polarized laser pulses and consists of a strong slowly varying field combined with a weak rapidly oscillating component. This setup allowed us to study, within a unified framework, the weak-field multiphoton regime, the strong-field tunneling regime, and the dynamically assisted Schwinger effect.

For the weak field alone, the momentum spectra exhibit the familiar shell structure of multiphoton pair production. For the strong field alone, the spectra are much smoother and admit a natural semiclassical interpretation in terms of pair creation near complex-time turning points followed by acceleration in the external field. When the two components are superimposed, the pair yield increases strongly, signaling dynamical assistance, while the spectra develop additional peaks and modulations due to the richer turning-point structure and the interference of different contributions.

The main focus of this work is helicity-resolved observables. For circular polarization, we found a clear helicity asymmetry: right- and left-handed electrons preferentially populate opposite momentum half-spaces. In the dynamically assisted regime, this asymmetry becomes more pronounced, showing that the weak fast component enhances not only the total yield but also the helicity selectivity of the process. Our central result is that the ratio of the momentum distributions for opposite helicities depends only weakly on the momentum magnitude and azimuthal angle and is governed predominantly by the polar angle with respect to the propagation axis. The corresponding asymmetry parameter increases with the weak-field frequency, thereby providing a compact quantitative measure of how dynamical assistance strengthens helicity effects. In particular, the present work identifies a simple organizing principle for helicity-resolved spectra in the dynamically assisted regime and establishes helicity asymmetry as a central observable of rotating strong-field backgrounds.

\acknowledgments

This study was funded by the Russian Science Foundation, project No.~24-72-10060.


\appendix*

\section{Locally-constant field approximation (LCFA)} \label{app:lcfa}

In this Appendix, we combine the semiclassical turning-point picture discussed in Sec.~\ref{subsec:ch2_spectra} with local pair-production rates and examine the performance of this approach for external fields with different polarizations.

\subsection{LCFA for circularly polarized fields} \label{subsubsec:lcfa_circ}

As discussed in Sec.~\ref{subsec:ch2_spectra}, semiclassical arguments already provide valuable qualitative information about the structure of the momentum spectra of the produced particles. A natural generalization of this picture is the locally constant field approximation (LCFA), where the spacetime dependence of the external field is taken into account locally and the corresponding contributions are then summed~\cite{kluger_prd_1992,aleksandrov_prd_2019_1,aleksandrov_kohlfuerst,sevostyanov_prd_2021,aleksandrov_symmetry,aleksandrov_sevostyanov_2025,tkachev_pra_2025}. A necessary condition for the applicability of the LCFA is a nonperturbative regime $(\gamma\ll 1)$; however, satisfying this condition does \emph{not} guarantee quantitative accuracy. Here we assess this approximation by comparing its predictions with the exact numerical results obtained from the QKE system.

Let us now discuss how the LCFA idea can be used to compute asymptotic momentum distributions and the total mean number of produced particles. Let $\mathbf{p}$ be the final kinetic momentum of a created electron. In the gauge $\mathbf{A}(+\infty)=0$, following the classical equation of motion, $\mathbf{q}(t)=\mathbf{p}-e\mathbf{A}(t)$, we can propagate the momentum backward in time. The observed electron with final momentum $\mathbf{p}$ is associated with a set of creation times $t_*^{(i)}$ corresponding to turning points determined by
\begin{equation}
q_{\parallel}\big(t_*^{(i)}\big)=0,
\label{eq:turning_points}
\end{equation}
where $q_{\parallel}$ denotes the component of the kinetic momentum parallel to the (instantaneous) electric field direction, and $i=1,2,\dots,M$ enumerates all such turning points contributing to the same final momentum $\mathbf{p}$.

Treating the external field as locally constant in the vicinity of each turning point, we evaluate
the pair-production probability using the standard constant-field expression~\cite{schwinger_1951,nikishov_constant},
\begin{equation}
W_i(\mathbf{p})=
\exp\!\left[
-\frac{\pi \pi_{\perp i}^2}{|e\mathbf{E}(t_*^{(i)})|}
\right],
\label{eq:LCFA_exp}
\end{equation}
where $\pi_{\perp i}^2 \equiv m^{2}+\mathbf{q}_{\perp}^{2}(t_*^{(i)})$ and $\mathbf{q}_{\perp}$ is the component of the kinetic momentum orthogonal to the field direction. The contributions from different turning points must be combined with the Pauli principle taken into account. This can be done via the iterative summation scheme~\cite{kluger_prd_1992,sevostyanov_prd_2021}
\begin{equation}
f^{(i)}(\mathbf{p})
=
f^{(i-1)}(\mathbf{p})
+\big[1-2f^{(i-1)}(\mathbf{p})\big]\,W_i(\mathbf{p}),
\label{eq:LCFA_f_i}
\end{equation}
where $i=1,2,\dots,M$ and $f^{(0)}(\mathbf{p})=0$. The spin-summed density is given by $f^{(e^-)}_{\text{LCFA}}(\mathbf{p}) = 2f^{(M)}(\mathbf{p})$.

For sufficiently small probabilities $W_i$, the total number of produced pairs in the LCFA may be estimated by directly summing the probabilities and including the spin degeneracy,
\begin{equation}
\frac{(2\pi)^{3}}{V}\,N_{\text{LCFA}} = 2\int d\mathbf{p}\,
\sum\limits_{i} W_i(\mathbf{p}).
\label{eq:LCFA_N_int}
\end{equation}
It is convenient to introduce an auxiliary integration over time,
\begin{align}
\frac{(2\pi)^{3}}{V}\,N_{\text{LCFA}} &= 2\int d\mathbf{p} \int dt \, \exp\!\left[
-\frac{\pi \pi_{\perp}^2}{|e\mathbf{E}(t)|}
\right] \nonumber \\
{}&\times \delta (q_\parallel (t)) |q_\parallel' (t)|,
\end{align}
and to perform a change of variables in the
momentum integral, $\{\mathbf{p}\}\to\{q_{\parallel},\mathbf{q}_{\perp}\}$. After integrating over $q_\parallel$, one obtains
\begin{align}
\frac{(2\pi)^{3}}{V}\,N_{\text{LCFA}} &= 2\int dt \int d\mathbf{q}_{\perp}\,
\exp\!\left[
-\frac{\pi\,(m^{2}+\mathbf{q}_{\perp}^{2})}{|e\mathbf{E}(t)|}
\right] \nonumber \\
{}&\times \bigg | 
\,|e \mathbf{E}(t)|
+
\frac{\mathbf{q}_{\perp}\cdot \dot{\mathbf{E}}(t)}{|\mathbf{E}(t)|}
\bigg |.
\label{eq:LCFA_N_our_res}
\end{align}
This result, derived for an arbitrary external-field polarization, is more general than the commonly used LCFA expression for the total number of pairs produced by a linearly polarized electric field (see, e.g., Ref.~\cite{sevostyanov_prd_2021}),
\begin{equation}
\frac{(2\pi)^{3}}{V}\,N_{\text{LCFA}}
=
2\int dt \, |e \mathbf{E}(t)|^{2}\,
\exp\!\left[
-\frac{\pi m^{2}}{|e \mathbf{E}(t)|}
\right].
\label{eq:LCFA_N_sev}
\end{equation}
If we discard the $\dot{\mathbf{E}}$ term in Eq.~\eqref{eq:LCFA_N_our_res} and integrate over $\mathbf{q}_\perp$, we recover Eq.~\eqref{eq:LCFA_N_sev}. The additional $\dot{\mathbf{E}}$ contribution that distinguishes \eqref{eq:LCFA_N_our_res} from
\eqref{eq:LCFA_N_sev} is related to the \emph{rotation} of the field direction, which affects the longitudinal ($q_\parallel$) volume of the occupied states. For linearly polarized fields, this contribution vanishes.

\begin{figure}[b]
  \centering \includegraphics[width=0.99\linewidth]{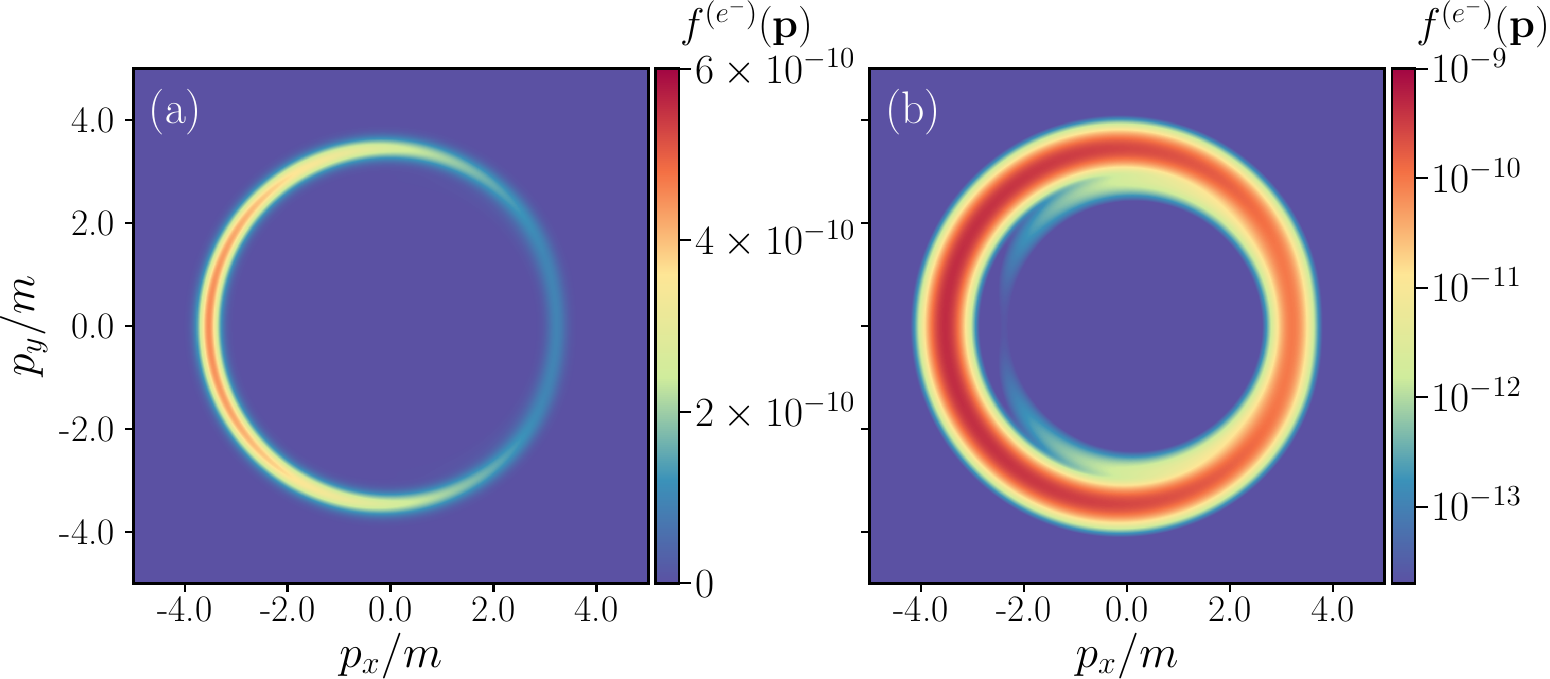}
\caption{Spin-summed electron momentum distributions computed within the LCFA for the \emph{strong-field-only} configuration described by the vector potential~\eqref{eq:field} ($E_1=0.2E_{\mathrm c}$, $E_2=0$, $\Omega=0.04m$, $\sigma=10$; circular polarization in the $xy$ plane). Panel (a): linear scale. Panel (b): logarithmic scale. The LCFA reproduces the overall spectral support of the exact QKE result, while underestimating the absolute densities (cf.~Fig.~\ref{fig:strong_only}).}
\label{fig:sum_strong_LCFA}
\end{figure}

In Fig.~\ref{fig:sum_strong_LCFA} we show the electron momentum distribution produced by the strong
field alone, computed within the LCFA (linear and logarithmic scales). These spectra are in qualitative agreement with the exact QKE results displayed in Fig.~\ref{fig:strong_only}. This is expected because the pair creation
mechanism for the strong field proceeds in the tunneling regime, which favors a better agreement
with the LCFA. Nevertheless, the predicted number densities are somewhat underestimated compared to the exact solution. This is not surprising, since the Keldysh parameter of the strong component is $\gamma_1 = 0.2$, which is not asymptotically small in this setup. More importantly, as shown in Ref.~\cite{sevostyanov_prd_2021}, the true criterion for the LCFA validity is $|eE_1|^{3/2}/\Omega \gg m^2$, whereas in the present case the left-hand side amounts only to $2.2$. Finally, we note that the LCFA completely neglects interference patterns, since Eq.~\eqref{eq:LCFA_f_i} is formulated at the level of probabilities rather than amplitudes.

\begin{figure}[t]
  \centering
      \includegraphics[width=\linewidth]{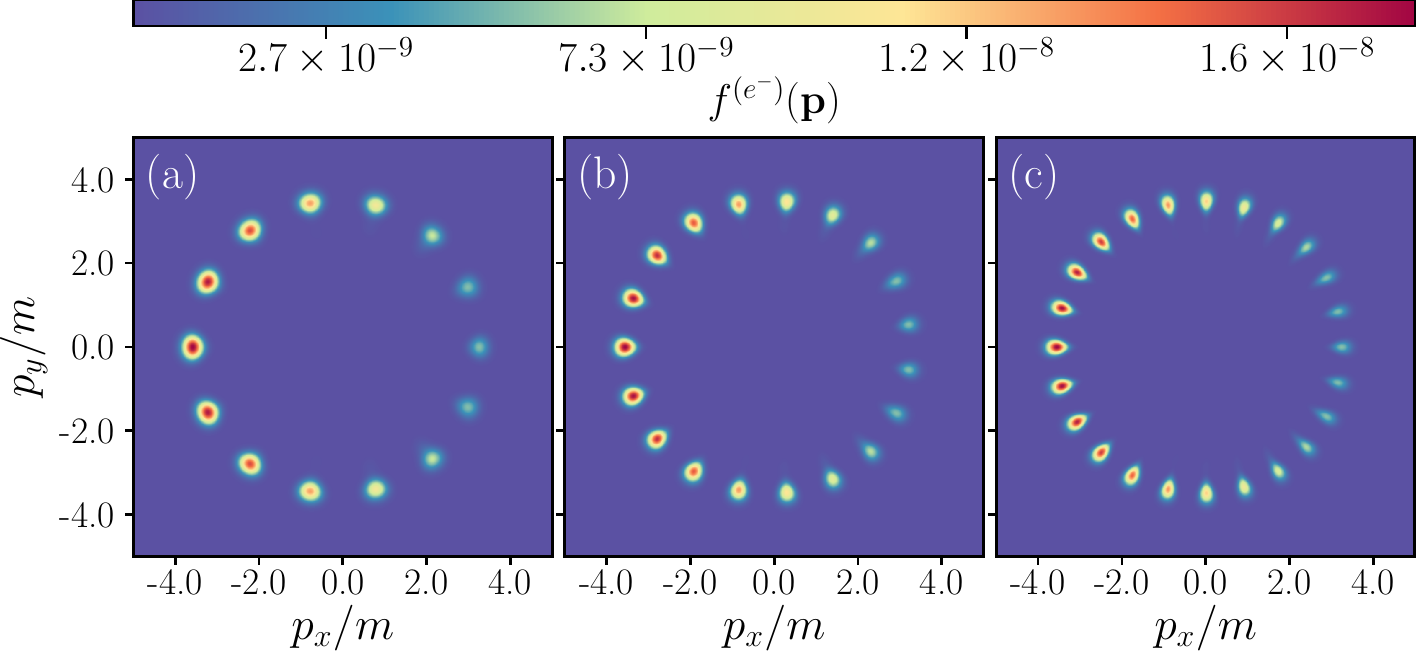}
  \caption{Spin-summed electron momentum distributions in the \emph{combined} strong and weak fields~\eqref{eq:field} computed within the LCFA.
  Panels: (a) $\omega=0.6m$, (b) $\omega=0.8m$, (c) $\omega=m$.
  The LCFA captures the qualitative structure of the exact spectra (Fig.~\ref{fig:combined_spectra}) but strongly underestimates the densities.}
  \label{fig:sum_comb_LCFA}
\end{figure}

\begin{figure}[t]
  \centering
      \includegraphics[width=0.95\linewidth]{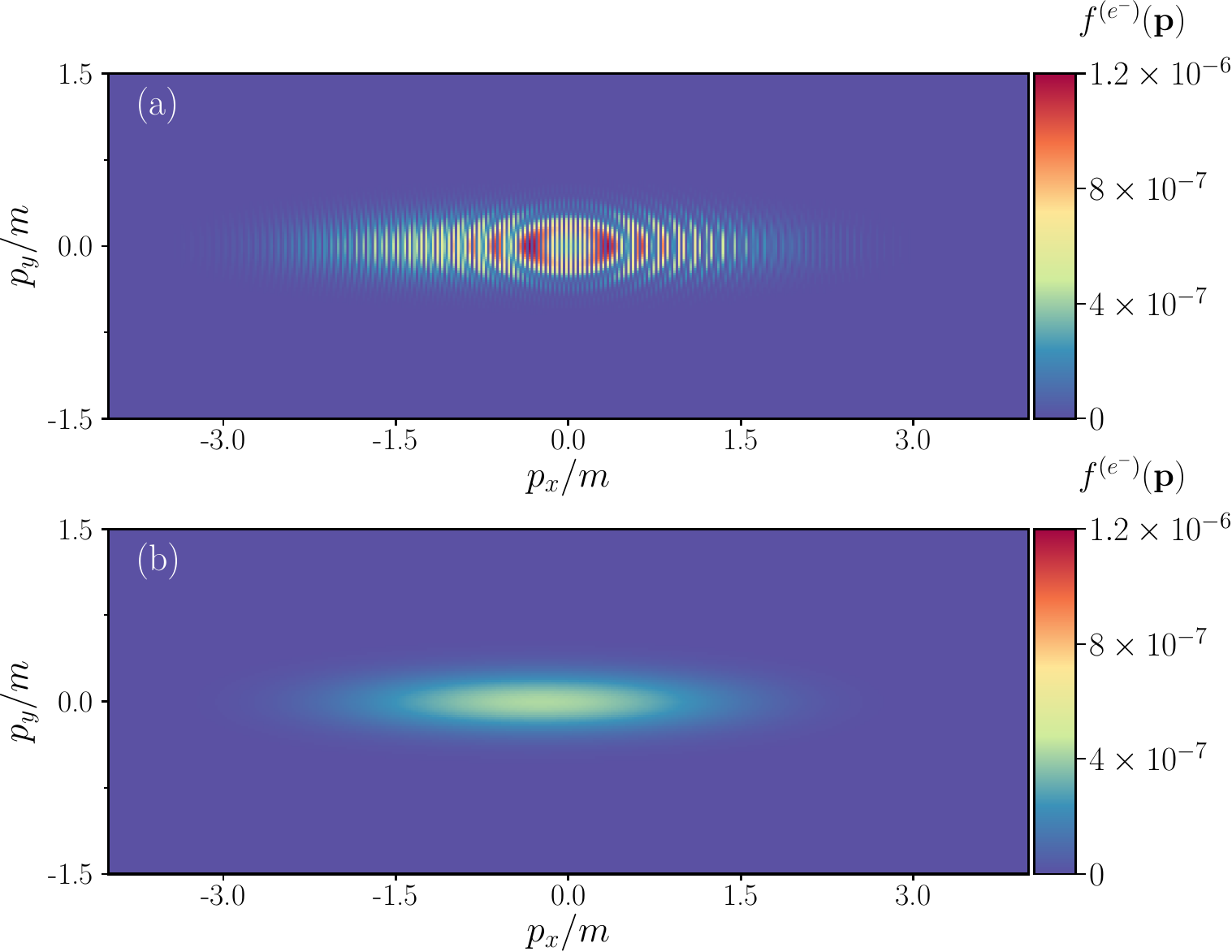}
\caption{Spin-summed electron momentum distributions produced by the \emph{strong field only} ($E_2=0$) for linear polarizations ($\delta_1=\delta_2=0$), shown in the $p_x p_y$
plane at $p_z=0$. Panel~(a): exact QKE result. Panel~(b): LCFA prediction.}
  \label{fig:sum_strong_LCFA_LP}
\end{figure}

We now apply the same approximation to the distributions obtained in the combined strong and weak fields. The LCFA spectra of the produced electrons are shown in Fig.~\ref{fig:sum_comb_LCFA} (to be compared with the exact results in Fig.~\ref{fig:combined_spectra}). The LCFA captures the qualitative structure of the spectra and may be viewed as a systematic extension of the semiclassical reasoning discussed in Sec.~\ref{subsec:ch2_spectra}. However, the LCFA does not capture the dynamical-assistance mechanism induced by the fast weak field. This is reflected in the comparatively weak dependence of the LCFA asymptotic distributions on the frequency of the weak component. As a result, the densities predicted by LCFA can be underestimated by several orders of magnitude compared to the exact spectra.

Thus, the LCFA is a useful tool for predicting the qualitative structure of the momentum spectra. Quantitatively reliable estimates are obtained essentially only for a single field acting in the nonperturbative regime and in the absence of pronounced interference effects. For dynamically assisted scenarios, for fields in or near the perturbative regime, and for helicity-resolved spectra, the QKE approach is the more reliable and versatile method.

\subsection{LCFA for linearly polarized fields}
\label{subsubsec:lcfa_linear}

We now briefly discuss linearly polarized fields, corresponding to $\delta_{1}=\delta_{2}=0$ in Eq.~\eqref{eq:field}. Figure~\ref{fig:sum_strong_LCFA_LP} shows the asymptotic
distribution of particles created by the strong field, computed both from the exact QKE solution
and within the LCFA. As noted above, the chosen field parameters do not fully justify the LCFA, so the underestimation of the densities is evident. Moreover, the LCFA fails to reproduce the complex structure of the spectrum
given by the exact QKE solution. This is explained by the strong influence of interference effects
on the asymptotic distribution in the case of a strong linearly polarized field (a large number of
pronounced interference features is also anticipated within the semiclassical picture discussed in Sec.~\ref{subsec:ch2_spectra}).

\begin{figure}[t]
  \centering
\includegraphics[width=0.95\columnwidth]{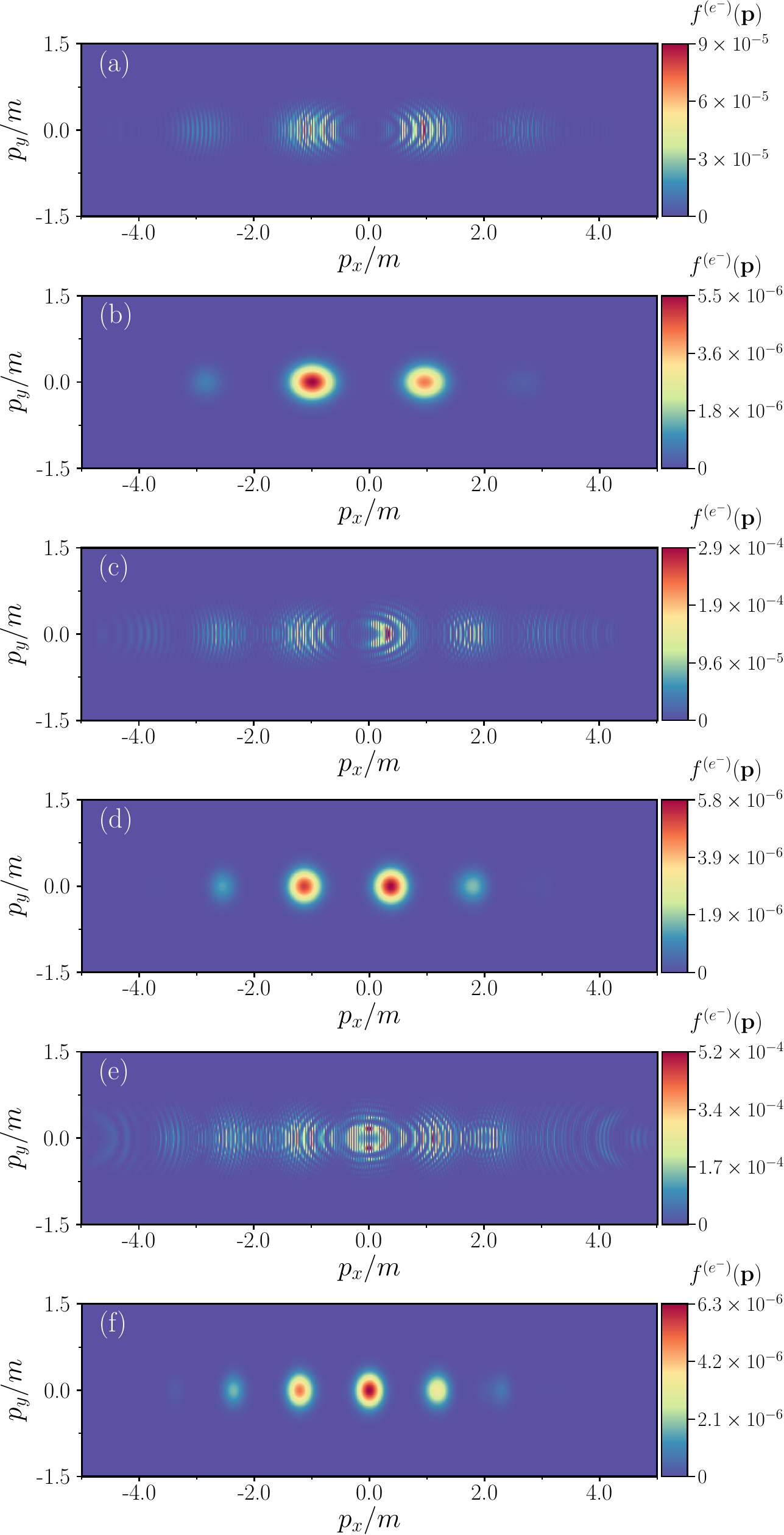}
\caption{Spin-summed momentum distributions in the \emph{combined} strong and weak fields~\eqref{eq:field} for $\delta_1 = \delta_2 = 0$, shown in the $p_x p_y$ plane at $p_z=0$. Panels (a), (c), and (e) correspond to the exact QKE results, while panels (b), (d), and (f) show the LCFA predictions. The weak-field frequency is $\omega=0.6m$ in panels (a) and (b), $\omega=0.8m$ in panels (c) and (d), and $\omega=m$ in panels (e) and (f). The comparison demonstrates that, although the LCFA reproduces the positions of the main maxima, it does not capture the pronounced interference structure present in the exact spectra and substantially underestimates the particle densities.}
\label{fig:LCFA_high}
\end{figure}

In Fig.~\ref{fig:LCFA_high} we compare the exact QKE spectra and the LCFA prediction for the combination of
the strong and weak fields in the linearly polarized configuration. Although the LCFA correctly identifies the regions of maximal intensity, strong interference effects
and the dynamical enhancement by the weak field do not allow this semiclassical approach to
quantitatively reproduce the exact distributions. We also note that, for linearly polarized fields,
the maximal particle density computed exactly is about an order of magnitude larger than for the
combination of circularly polarized fields.

Therefore, when applying the LCFA to asymptotic momentum distributions, one must assess how important interference effects are for the configuration under consideration, even for strong fields in the tunneling regime. At the same time, LCFA can provide useful qualitative information on the
spectral structure. The QKE method, in turn, contains no semiclassical assumptions and yields accurate momentum-resolved particle densities. Moreover, the QKE approach allows one to analyze spectra at fixed helicity.


\end{document}